\documentstyle[11pt,epsf,aaspp4,flushrt]{article}

\newcommand{\Lya}{Ly$\alpha\ $}

\newcommand{\LCDM}{$\Lambda$CDM\ }
\newcommand{\HI}{\hbox{H~$\rm \scriptstyle I\ $}}
\newcommand{\HII}{\hbox{H~$\rm \scriptstyle II\ $}}
\newcommand{\HeI}{\hbox{He~$\rm \scriptstyle I\ $}}
\newcommand{\HeII}{\hbox{He~$\rm \scriptstyle II\ $}}
\newcommand{\HeIII}{\hbox{He~$\rm \scriptstyle III\ $}}
\newcommand{\NHI}{N_{\rm HI}}
\newcommand{\cms}{\,{\rm cm$^{-2}$}\,}
\newcommand{\kms}{\,{\rm km\,s$^{-1}$}\,}

\newcommand{\etal}{{ et~al.~}}

\begin{document}


\title{Hydrodynamical Simulations of the Lyman Alpha Forest:\ Model
Comparisons \\}

\author{Marie E. Machacek}
\affil{Physics Department, Northeastern University, Boston, MA 02115}

\author{Greg L. Bryan\altaffilmark{1}}
\affil{Physics Department, Massachusetts Institute of Technology,
                Cambridge, MA 02139}

\author{Avery Meiksin}
\affil{Institute for Astronomy, University of Edinburgh,
      Royal Observatory, Blackford Hill, Edinburgh\ EH9\ 3HJ}

\author{Peter Anninos}
\affil{University of California, Lawrence Livermore National Laboratory,
         Livermore, CA 94550 }

\author{Daniel Thayer, Michael Norman\altaffilmark{2}}
\affil{Astronomy Department, University of Illinois at Urbana-Champaign, 
         1002 West Green Street, Urbana, IL 61801}

\and

\author{Yu Zhang}
\affil{Prowess Systems, 1370 Ridgewood Dr., Ste. 20, Chico, CA 95973}

\altaffiltext{1}{Hubble Fellow}
\altaffiltext{2}{also at Laboratory for Computational Astrophysics,
                National Center for Supercomputing Applications,
                University of Illinois at Urbana-Champaign,
                405 Mathews Ave., Urbana, IL 61801 }


\begin{abstract}

We investigate the properties of the Lyman alpha forest as predicted
by numerical simulations for a range of currently viable cosmological
models.  This is done in order to understand the dependencies of the
forest on cosmological parameters.  Focusing on the redshift range
from two to four, we show that: (1) most of the evolution in the
distributions of optical depth, flux and column density can be
understood by simple scaling relations, (2) the shape of optical depth
distribution is a sensitive probe of the amplitude of density
fluctuations on scales of a few hundred kpc, (3) the mean of the $b$
distribution (a measure of the width of the absorption lines) is also
very sensitive to fluctuations on these scales, and decreases as they
increase.  We perform a preliminary comparison to observations, where
available.  A number of other properties are also examined, including
the evolution in the number of lines, the two-point flux distribution
and the HeII opacity.
\end{abstract}



\section{Introduction}
\label{sec:introduction}

Several numerical simulations of the \Lya forest 
in cold dark matter (CDM) dominated cosmologies
have been performed in recent years and compared with observations 
(Cen \etal 1994; Zhang, Anninos, \& Norman 1995; Hernquist \etal 1996;
Miralda--Escud\'e \etal 1996; Zhang \etal 1996;  Dav\'e \etal 1997; Bond \&
Wadsley 1997; Zhang \etal 1998). Remarkably all the
simulations have been able to reproduce the measured neutral hydrogen column
density distribution, the size of the absorbers (Charlton \etal 1997),
and the line number evolution reasonably well,
despite the differences in the cosmological models used:\ Cen \etal adopt a
$\Lambda$CDM model; Zhang \etal investigate sCDM models with both an unbiased
and a cluster scale normalization; Hernquist \etal (1996) evolve an sCDM model
with a cluster scale normalization. The distribution of Doppler parameters
has fared somewhat less well:\ the predicted distribution peaks toward lower
values than observed when the simulations are performed with adequate
resolution (Bryan \etal 1998; Theuns \etal 1998). Nonetheless, the generally
good agreement with observations of the \Lya forest suggests that the models
are capturing the essential physical properties of the absorbers.  This has 
prompted recent work by Croft \etal (1998) aimed at using  flux 
statistics of the observational data to extract the fluctuation 
spectrum of the underlying cosmology. We are thus encouraged to investigate 
the possibility that differences in the statistical properties of the \Lya 
forest predicted by different cosmological models may provide a means of 
testing the models.

The objective of this paper is to compare the \Lya forest statistics
derived from simulations in different cosmological models and to investigate
what key properties of the cosmological models control a given 
statistic. We present results from nine numerical simulations 
using five different background cosmological models, three of which are 
flat with no cosmological constant, one is open, and one is flat with a 
nonzero cosmological constant. For five of the simulations, which we 
will refer to as the model comparison study, the parameters of the 
cosmological models have been selected by their ability to match the local 
or low redshift observations, although all of these models except the 
standard cold dark matter (sCDM) model are also consistent with COBE 
measurements of the cosmic microwave background.  A tilted CDM model is 
further designed to match COBE constraints on the normalization of the power 
spectrum  on large scales.  In the remaining four simulations we keep the 
underlying cosmology fixed (sCDM) 
while varying the normalization of the fluctuation power 
spectrum in order to clarify the dependence of the \Lya statistics on this 
parameter.  The radiation field is normalized to the absorption properties of 
the \Lya forest as measured at high redshifts.  While the emphasis in this 
paper is on comparing cosmological 
models, we also test how well the models are doing using the tabulated 
statistics of the \Lya forest as determined primarily by Kim \etal (1997) 
for several QSO lines-of-sight. A more complete comparison with existing 
data from several observational groups will be presented in Meiksin \etal 
(1999).  

The paper is organized in the following way. In \S\ref{sec:models} we 
describe the cosmological models and simulation technique.  
In \S\ref{sec:rawopacity} 
we investigate the model differences, power dependence, and redshift 
evolution in the raw opacity data as characterized by nonparametric 
statistics of the flux and optical depth.  In \S\ref{sec:spectrum} we 
present a line analysis of the spectra generated by the various simulations 
focusing on the column density distribution and line number evolution 
statistics. In \S\ref{sec:bparm} we discuss the Doppler b parameter 
distributions and related nonparametric statistics, and in 
\S\ref{sec:helium} we  present model predictions for \HeII absorption.  We 
summarize our results in \S\ref{sec:summary}.


\section{The Models and Simulations}
\label{sec:models}

All the model background spacetimes we consider are in the context of
Cold Dark Matter (CDM) dominated cosmologies. We examine the following
five models: a standard critical density flat CDM model (sCDM),
a flat CDM model with a nonvanishing cosmological constant (\LCDM), a
topologically open CDM model (OCDM), the standard CDM model but with
the power spectrum of the density perturbations tilted (tCDM) to match the 
normalization on large scales as determined from the COBE measurements of 
the Cosmic Microwave Background (Bunn \& White 1997), and a flat critical 
density mixed dark matter model with a hot component added to the CDM (CHDM).
There are several important and well--established astrophysical measurements 
which constrain the various combinations of cosmological parameters.
The parameters for each model, which we list in Table \ref{tab:sim_par}, 
have been determined to provide good consistency with these observations.
For example, the combination $\Omega_B h^2$ is restricted by
Big Bang nucleosynthesis constraints and the measured abundance
of primordial deuterium to lie in the range $0.015$--$0.025$
(Copi, Schramm, \& Turner 1995; Burles \& Tytler 1998).
In addition, because the \HI column density scales approximately
as $(\Omega_bh^2)^2$ for a fixed UV radiation intensity, we choose 
$\Omega_b$ and $h$ so that $\Omega_bh^2$ is the same for three of the
models sCDM, \LCDM and OCDM. The fluctuation normalization in a sphere of 
$8 h^{-1}$ Mpc is defined to match observations of the number density of 
galaxy clusters (White, Efstathiou, \& Frenk 1993; Bond \& Myers 1996) in 
all the models. In addition, a tilt has been applied to the CDM power 
spectrum in the tCDM model in order to approximately match the amplitude of 
the CMB quadrupole as measured by COBE (Bunn \& White 1997). The 
cosmological constant in the \LCDM case is consistent with the upper limit
($\Omega_\Lambda<0.7$) of Maoz and Rix (1993) and the best fit parameters of 
Ostriker and Steinhardt (1995). One of the major problems with the sCDM model
is its difficulty in matching observations of the large scale structures in 
the universe. Since the standard CDM model is historically one of the most 
studied models, however, we use it as our canonical model 
to which the perhaps more viable additional models considered here may
be compared and through which we investigate the dependence of the \Lya 
statistics on the fluctuation power spectrum.  We refer the reader to 
Zhang \etal (1995, 1997, 1998) for further details and results from our 
previous sCDM simulations.

\begin{figure}
\epsfxsize=3.5in
\centerline{\epsfbox{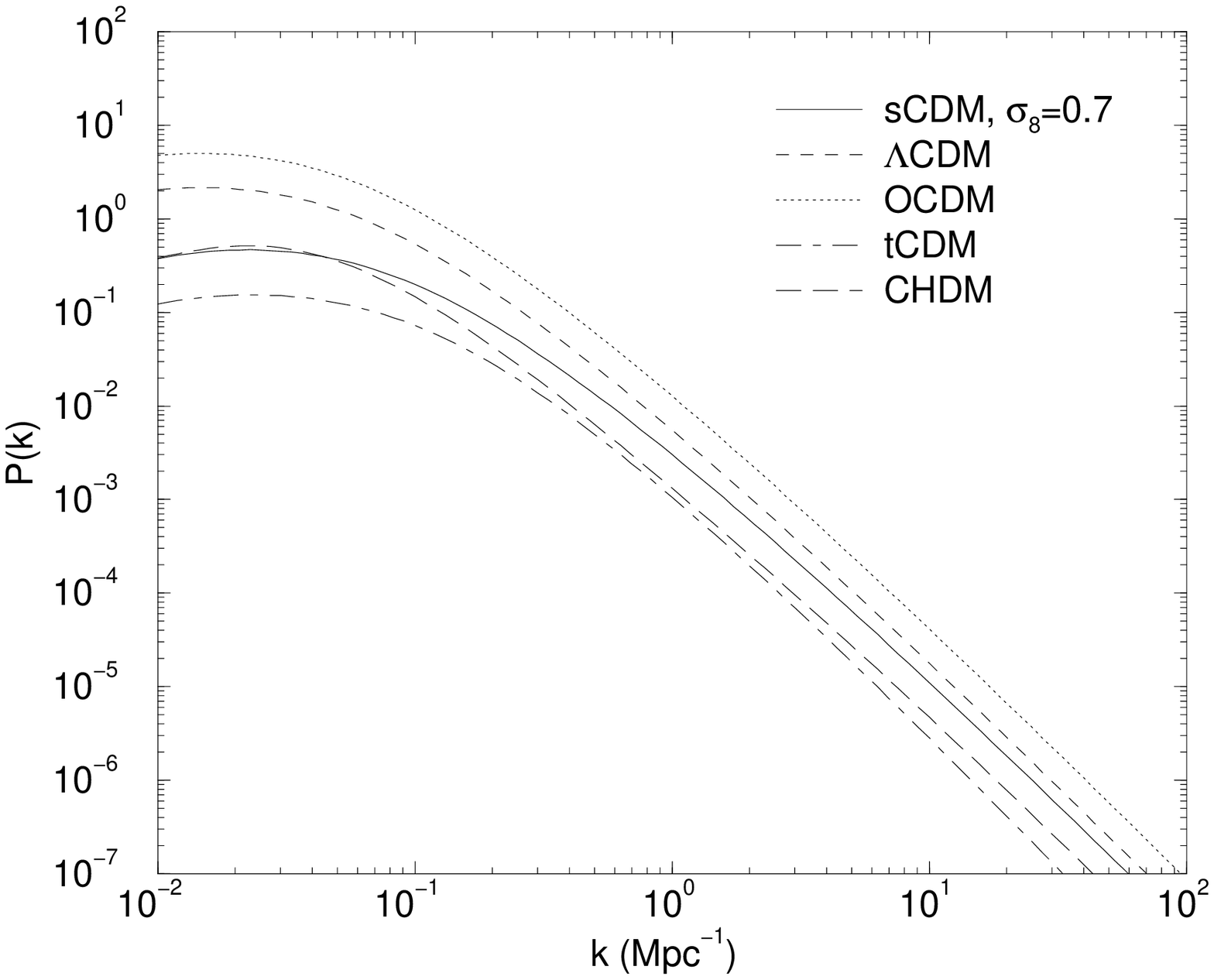}}
\caption{
Power spectra for the models in the model comparison study (sCDM 
$\sigma_8=0.7$, \LCDM, OCDM, tCDM, and CHDM) plotted for a linear evolution 
to $z=3$. }
\label{fig:powspec}
\end{figure}

 The initial data were generated using COSMICS (Bertschinger 1995) with the 
BBKS transfer function (Bardeen \etal 1986) to compute 
the starting redshifts and the initial particle positions and 
velocity perturbations appropriate for all models except CHDM. We used 
CMBFAST (Seljak \& Zaldarriaga 1996) to solve the linearized 
Boltzmann equations to set the initial conditions for CHDM. For the comoving 
box size adopted (9.6 Mpc) and the corresponding comoving grid cell size 
(37.5 kpc in our high resolution runs), the relevant wavenumber domain of 
the simulations at $z=0$ is $168 > k > 0.65$~Mpc$^{-1}$, where $k=2\pi/\ell$
and $\ell$ is the length scale. Over this domain, the
sCDM, \LCDM and OCDM models all have a similar power distribution. 
The tCDM and CHDM models, on the other hand, have an overall lower 
normalization (see Table \ref{tab:sim_par}) in addition to a steeper slope
that drops slightly more sharply than the other models over the smaller 
scales.  In Figure \ref{fig:powspec} we show the linear power spectra for 
these models evolved to $z=3$, the redshift at which we present many of our 
results. Since previous work (Zhang \etal 1998) indicates that the sizes of 
the low column density absorbers at $z \sim 3$ are $\sim 100$ kpc, it is 
useful to characterize the models in terms of their power at these small 
fluctuation scales.  A useful measure of this power, introduced by Gnedin 
(1998), is
\begin{equation}
\sigma^2_{34} = \int_{0}^{\infty} P(k) e^{-2k^2/k^2_{34}} 
{{k^2 dk} \over {2 \pi^2}}
\label{eq:sig34}
\end{equation}
(where $k=2\pi /\ell$, $P(k)$ is the linear power spectrum at $z=3$ and 
$k_{34} = 34 \Omega_0^{1/2} h$ Mpc$^{-1}$). This is also listed for each model
in Table \ref{tab:sim_par}.  
   
In addition to specifying a cosmological model,
it is also necessary to include a background UV radiation field
to ionize the IGM. We have implemented the spectrum
computed by Haardt \& Madau (1996) for a flat universe on the basis
of radiation transfer in a clumpy universe and the measured
luminosity function of QSOs, accounting for QSO
sources, absorption by the \Lya forest and Lyman limit systems, and 
re-emission of the recombination radiation from the absorbing clouds.
This spectrum reionizes the universe between redshifts $7$ and $6$ and peaks 
at about $z=2$. Because the OCDM model corresponds to a cosmology not 
considered by Haardt \& Madau (1996), we use the field for a flat universe 
for this model also, noting that the neutral fractions are in any case 
rescaled, as described below, to match observations.  We also 
note that only clouds which are optically thin at the Lyman edge 
are considered in this paper. Hence the optically thin limit is a good 
approximation and it is not necessary to account for self--shielding and 
radiative transfer of the external ionizing radiation field.

The numerical computations were performed using two different
numerical codes, Kronos and Hercules, each in a simulation box of length 
$9.6\rm\, Mpc$ comoving with the universal expansion.  Kronos (Bryan \etal 
1995) is a single grid Eulerian code that uses a particle-mesh algorithm to 
follow the dark matter and the piecewise parabolic method (PPM) to simulate 
the gas dynamics.  Since non--equilibrium chemistry and cooling processes 
can be important, six particle species (\HI, \HII, \HeI, \HeII, \HeIII and 
the electron density) are followed with a sub-stepped backward 
finite-difference technique (Abel \etal 1997; Anninos \etal 1997).  This is 
the same non--equilibrium chemistry and cooling model used in our previous 
studies of the \Lya Forest (Zhang \etal 1995, 1997, 1998; 
Charlton \etal 1997; Bryan \etal 1998). For sCDM, \LCDM, OCDM, and tCDM we 
use $256^3$ grid cells in the 
simulation box to follow the evolution of $128^3$ dark matter particles. This 
results in a spatial resolution of $\Delta x=37.5\rm\, kpc$. For 
CHDM $128^3$ grid cells are used with $64^3 (128^3)$ cold (hot) dark matter 
particles, respectively, resulting in a lower spatial 
resolution,  $\Delta x = 75 \rm\, kpc$, for this model.  For the sCDM 
simulations with varying $\sigma_{8h^{-1}}$ we present simulations with both 
$128^3$ and $256^3$ grid cells resulting in both low and high spatial 
resolutions of $75\rm\, kpc$ and $37.5\rm\, kpc$, respectively. 

We have also simulated three of our models (sCDM, \LCDM, OCDM) with a 
different numerical code, Hercules (Anninos, Norman \& Clarke 1994; Anninos 
\etal 1997).  Hercules is a nested grid code that utilizes a multiscale PM 
method for the dark matter, artificial viscosity methods for the
baryonic fluid, and the same non--equilibrium chemistry and cooling model as 
above. The simulations produced from this code use $128^3$ particles and 
$128^3$ cells for both the nested and parent grids. However, in order to 
derive a more representative sample for statistics, the results discussed in 
this paper are extracted from the parent grid only. Thus these simulations
are of lower spatial resolution than most of the Kronos simulations, although
the dark matter mass resolution is the same.  For statistics that are 
insensitive to spatial resolution, a comparison of the results of the two 
codes is useful to insure that simulation results are robust against 
changes in numerical technique.  

Synthetic spectra are generated along $300 (900)$ random lines of sight 
through the Kronos (Hercules) simulated volume using the method of 
Zhang \etal (1997) including the effects of peculiar velocity and thermal 
broadening of the gas. (We have verified that decreasing the sample size from
900 to 300 for the Kronos data does not affect the results except for a slight
increase in the scatter of the line properties for the highest,
optically thick column density systems, a regime where our results  
become unreliable anyway because of the absence of radiative transfer in the 
codes.) Since we are primarily concerned in this paper with a comparison of 
model predictions, we have not included noise or continuum fitting in the 
analysis. Furthermore the resolution of the spectra, $1.2\rm\, km/s$, is 
the same for all the simulations, a value that is smaller than current 
observations. However, we have shown elsewhere (Bryan \etal 1998) 
that, as long as we restrict ourselves to high quality observational data, 
the impact of not including these observational difficulties is small. In 
addition to analyzing the raw optical depth and flux distributions, line 
lists are extracted from the data using a Voigt profile fitting procedure. 
This is described in more detail elsewhere (Zhang \etal 1997), but we 
outline it briefly here.  First, maxima in 
the optical depth distribution are identified as line centers. Then Voigt 
profiles are fit, using a non-linear minimization, to the part of the 
spectrum which is above $\tau_{HI} = 0.05$ and between neighbouring minima.
This results in the same spectral threshold $F_t\equiv e^{-\tau_t}=0.95$ as 
the high resolution Keck HIRES spectrometer.   Each line of sight 
chosen produces a sample spectrum with on the order of  $10$--$100$ lines 
per redshift interval $\delta z = 0.1$ depending on the redshift and 
cosmological model.  The statistics of these linelists are discussed in 
\S\ref{sec:spectrum} and \S\ref{sec:bparm}.

\begin{figure}
\epsfxsize=3.5in
\centerline{\epsfbox{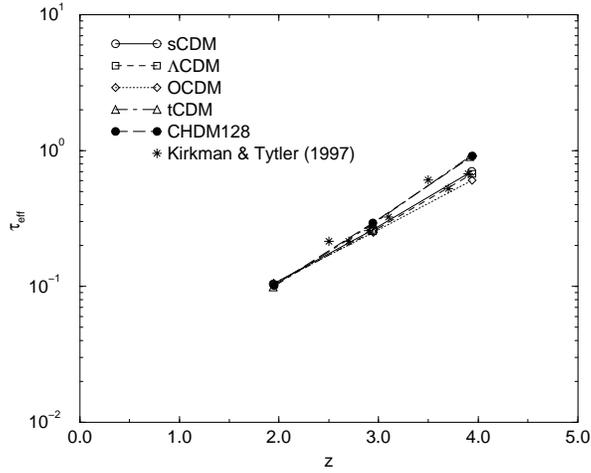}}
\caption{
Effective HI optical depth as a function of redshift for simulations of the 
model comparison study normalized to $\bar \tau=0.3$ at $z=3$ compared to 
recent data by Kirkman \& Tytler 1997 (denoted by stars).}
\label{fig:taunorm}
\end{figure}

The {\it amplitudes} of the distributions found in the models cannot be used
as a basis for comparing the models since they may be arbitrarily
re-scaled for any individual model using the ionization bias factor
$b_{\rm ion}=\Omega_B^2/\Gamma$, where $\Omega_B$ is the fraction of the 
critical density carried in baryons and $\Gamma$ is the Haardt-Madau (1996) 
parameterization of the metagalactic UV ionizing background extracted from 
the observed distribution of quasars.
It is important to normalize all the models consistently before
comparing the shapes of any of the distributions. This may be done in a
variety of ways. We do so by matching the mean \HI opacity in each simulation
to the measured intergalactic \HI opacity at $z=3$. In Zhang \etal (1997),
we found the opacity measurements of Steidel \& Sargent (1987) and Zuo \& Lin
(1993) gave a mean \HI opacity at $z=3$ of $\bar\tau_\alpha=0.27-0.35$,
although values as much as 30--60\% larger have been claimed 
(Press \etal 1993; Rauch \etal 1997). Because of the uncertainty in this 
measurement, we also require consistency with the number density of lines 
observed above a threshold of $\log\NHI=13.5$. Using the three quasars in 
Hu \etal (1995) for which lines in the full redshift range $3<z<3.1$ are 
listed, we find a total of 61 lines for the three lines--of--sight in this 
redshift interval with $\log\NHI>13.5$, for which the line lists should be 
complete. (An estimate based on using the available lines for all four QSO 
line lists in Hu \etal in the redshift interval $2.9<z<3.1$ gave essentially 
the same line density.) Normalized to $\bar\tau_\alpha=0.30$, the CHDM, 
sCDM, \LCDM, OCDM, and tCDM models predict, respectively, 60.8, 62.1, 62.7, 
63.7, and 59.5 lines, in close agreement with the observed number. 
Normalizing to $\bar\tau_\alpha=0.35$, the respective numbers of predicted 
lines are 73.7, 74.3, 73.9, 75.2, and 72.8. While these are not badly 
inconsistent with the observed number, they are all fairly high. We 
normalize the spectra according to $\bar\tau_\alpha=0.30$
throughout this paper, noting that this value is still not well agreed upon.
In Figure \ref{fig:taunorm} we plot a related statistic, $\tau_{eff}$ 
(Zhang \etal 1997) for the normalized spectra of our models and compare to 
recent data by Kirkman \& Tytler (1997). After normalization all of our 
models are consistent with the data over the redshift range $2 \le z \le 4$ 
considered by this paper. 


\section{Direct Optical Depth and Flux Measurements}
\label{sec:rawopacity}

  Historically \Lya absorption spectra have been analyzed in terms of the 
statistics of spectral line features and as such have been plagued with 
difficulties of the line fitting procedure such as line identification 
and blending. Many of these difficulties become increasingly severe at higher 
redshifts making the results of the analysis uncertain. It is thus natural 
to ask whether statistics dependent directly on the observed flux and optical
 depth without recourse to line fitting might be of use in describing the 
forest and discriminating among competing models. Statistics of this kind 
have recently been proposed by several authors (Miralda--Escud\'e \etal 1997;
Rauch \etal 1997; Cen 1997). Since these nonparametric measures are also 
easier to relate theoretically to the physical state of the absorbing gas, 
we begin our discussion with them.

\subsection{Optical Depth Probability Distribution Function}
\label{sec:tau}

\begin{figure}
\epsfxsize=3.5in
\centerline{\epsfbox{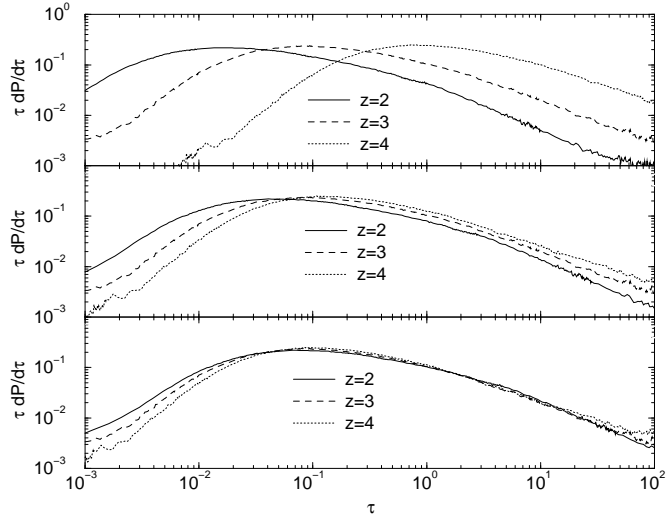}}
\caption{
(top) Optical depth probability distribution function for the high spatial 
resolution sCDM model with $\sigma_{8}=0.7$ for redshifts $z = 2,3$ and $4$.
(middle) The same optical depth probability distribution functions with 
$\tau$ scaled by $f(3)/f(z)$ where $f(z)=(1+z)^{4.8}/(\Gamma (z)H)$ is the 
naive scaling predicted assuming no redshift evolution of the gas overdensity.
(bottom) The same distributions with additional scaling due to the $z$ 
evolution of the peak of the density distribution shown in 
Figure \protect\ref{fig:overdens}.}
\label{fig:scdmtau}
\end{figure}

    The optical depth $\tau$ is related to the transmitted flux $F$ by 
$F=exp(-\tau)$. We define the optical depth probability distribution  
$dP/d\tau$ as the probability that a pixel will have optical depth 
between $\tau$ and $\tau + d\tau$. In Figure \ref{fig:scdmtau} we use 
spectra generated from 
the sCDM high resolution simulation to show  $\tau dP/d\tau$ versus $\tau$ 
for redshifts $z=2,3$ and $4$ (top panel). Although the peak of the 
distribution decreases and the distribution broadens slightly with 
decreasing redshift, 
the principal contributor to the redshift evolution seen in Figure 
\ref{fig:scdmtau} is the evolution of the optical depth $\tau$.  
Hui, Gnedin \& Zhang (1997) discuss in detail the dependence of $\tau$ 
on the distribution and properties of neutral hydrogen along the 
line of sight in an expanding universe. Since we would like to understand 
the redshift evolution of the optical depth in terms of simple scaling laws, 
we repeat some of their discussion 
here in order to the isolate the key factors controlling this redshift 
evolution and clarify the scaling law assumptions.  The optical depth is 
defined as 
\begin{equation}
\tau(\nu_o) = \int_{x_a}^{x_b} n_{HI}\sigma_\alpha\, {dx \over 1+z}
\label{eq:taudefx}
\end{equation}
where $\nu_o$ is the observed frequency, $n_{HI}$ is the number 
density of 
neutral hydrogen, $z$ is the redshift of the absorbing gas, $\sigma_\alpha$
is the absorption cross section for \Lya , and the integral is over the line 
of sight between the quasar ($x_a$) and the observer ($x_b$) in comoving 
coordinates. In practice the form of the \Lya absorption cross section limits
the integration range per absorber to a small portion of the line of sight.  
It is thus useful to make a change of variable to velocity coordinates $u$ 
about some characteristic average redshift $\bar z$ in the problem.  For 
example, for simulated data the redshift $\bar z$ might be a given output 
redshift for the simulation.  The observed frequency $\nu_o$ and the frequency
$\nu$ of the radiation in the absorber rest frame are then related by 
\begin{equation}
\nu = \nu_o(1+\bar z)(1+ u/c)
\end{equation}
where
\begin{equation}
u \equiv {H(\bar z)(x - \bar x) \over 1 + \bar z} + v_{pec}(x).
\label{eq:u}
\end{equation} 
$\bar x$ is the comoving position along the line of sight whose redshift is 
exactly $\bar z$, $v_{pec}$ is the physical velocity of the gas, and 
$H(\bar z)$ is the Hubble parameter defined by
\begin{equation}
H(\bar z) = H_0 
\sqrt{\Omega_m (1+\bar z)^3+(1-\Omega_m-\Omega_\Lambda)(1+\bar z)^2 + \Omega_\Lambda}.
\end{equation}
The first term in Equation \ref{eq:u} represents the contribution of the 
residual Hubble flow about the mean while the second term is due to the
physical bulk flow of the gas.  We assume $u/c << 1$ and neglect 
contributions from turbulent flows since they would be unlikely in the low 
column density regions we are considering.  Under this change of variable
the \Lya cross section becomes
\begin{equation}
\sigma_\alpha = {\sigma_{\alpha 0}c  \over b \sqrt{\pi}}e^{-(u - u_0)^2/b^2}
\end{equation}
where $\sigma_{\alpha 0}=4.5 \times 10^{-18}$ cm$^2$ sets the scale 
of the absorption cross section in terms of fundamental constants, $u_0$ is 
the velocity $u$ for which the frequency $\nu$ in the rest frame of the 
absorbing gas is equal to the \Lya frequency $\nu_\alpha$ and 
$b = \sqrt{2k_BT/m_p}$ is the thermal width.  For absorption lines of neutral
hydrogen with column densities $\NHI<10^{17}$ \cms the thermal profile 
dominates the cross section so we neglect the contribution of the natural 
line width to $\sigma_\alpha$.  The optical depth $\tau$ can now be 
written as  
\begin{equation}   
\tau = {\sigma_{\alpha  0} c \over \sqrt{\pi}}\sum_{streams} 
\int {n_{HI} \over b(1+\bar z)} 
\left|{du \over dx}\right|^{-1} e^{-(u - u_0)^2/b^2}\,du
\label{eq:taudef}
\end{equation}
The sum over streams represents the possibility that a given velocity $u$ 
corresponds to more than one position $x$.  Although the integration formally
runs over the full line of sight from quasar to observer, the
Gaussian form for the cross section effectively limits the $u$ integration to
a narrow range around $u_0$ (thus justifying our replacement of $z$ everywhere
by $\bar z$.)  To simplify notation we drop the bar letting $z$ represent 
$\bar z$ in what follows. 

\begin{figure}
\epsfxsize=3.5in
\centerline{\epsfbox{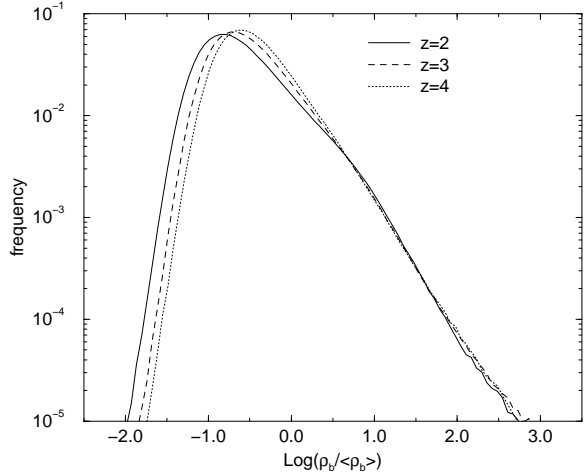}}
\caption{
Redshift evolution of the gas overdensity for the high spatial resolution 
sCDM model with $\sigma_{8}=0.7$.}
\label{fig:overdens}
\end{figure}

 We assume that the number density of hydrogen traces the baryon gas density
well. (There has been little metal production in these low density regions 
and there is no interaction that would cause the helium and hydrogen to 
separate.) Thus the number density of neutral hydrogen is 
$n_{HI}= \rho_b X_{HI}$ 
where $X_{HI}$ is the neutral fraction and $\rho_b$ the gas density. 
In ionization equilibrium 
(which is well satisfied except for the period of initial reionization)
the neutral fraction of hydrogen is  $X_{HI} \propto \rho_b T^{-0.7}$
such that the number density of neutral hydrogen (relevant to \Lya 
absorption) scales as   
\begin{equation}
n_{HI} \propto (\Omega_b h^2)^2 \Gamma^{-1}(z)(1+z)^6(1+\delta_b)^2T^{-0.7}
\end{equation} 
where $\delta_b$ is the baryon overdensity. Studies (Hui \& Gnedin, 1996; 
Weinberg \etal, 1996) of the equation of state for the gas 
find that for unshocked gas at low to moderate baryon overdensities 
($\delta_b \le 5$) the equation of state is well fit by a power law:
\begin{equation}  
T \propto (1+z)^{1.7}(1+\delta_b)^{\gamma -1}.
\end{equation}
  Thus
\begin{equation}
n_{HI} \propto (\Omega_b h^2)^2 \Gamma^{-1}(z)(1+z)^{4.8}
(1+\delta_b)^{2.7-0.7\gamma}.
\label{eq:nHI}
\end{equation}
For a uniform radiation field and reionization that occurs before $z=5$, as 
is the case in our simulations, $\gamma \approx 1.4$. This is in agreement 
with the value $\gamma\approx1.5$ found by Zhang et al. (1998) for clouds 
with column densities in the range $12.5<\log \NHI < 14.5$. 
Furthermore the assumption that most of the optical
depth arises from low column density absorbers, large structures 
whose overdensities and peculiar velocities are slowly varying compared to 
the thermal profiles, means that multiple streaming is rare, the sum over
streams in Equation \ref{eq:taudef} can be dropped, and 
$\left|{du \over dx}\right| \approx {H \over 1+z}$.  We then integrate 
over the thermal profile to obtain (Croft \etal 1997)
\begin{equation} 
\tau \propto {c\sigma_{\alpha 0}(\Omega_B h^2)^2 \over 
\Gamma(z)H} (1+z)^{4.8}(1+\delta_{b})^{1.7}.
\label{eq:taufull}
\end{equation}
Note that in this limit $\tau$ need no longer have a thermal profile about
its maximum (Hui, Gnedin \& Zhang, 1997).
If $\delta_b$ is evolving slowly over this redshift range, $\tau$ should 
scale as 
\begin{equation}
\tau \propto {(1+z)^{4.8} \over \Gamma (z)H }.
\label{eq:tausim}
\end{equation}  
In the middle panel of Figure \ref{fig:scdmtau} we use this simple scaling 
law to rescale the $z=2$ and $z=4$ sCDM distributions from the top panel 
 to $z=3$,
the redshift at which all the models are normalized.  We do this in order to 
test how well the simulations obey this simple scaling relation: if they 
followed it exactly then all three curves would overlap.
Most, but not all, of the redshift evolution of this distribution is 
accounted for by the scaling of $\tau$ given in Equation \ref{eq:tausim}.
Since the evolution of the metagalactic UV radiation field $\Gamma$ is 
relatively slight 
over this redshift range, we are left with the remarkable conclusion that 
{\it most of the evolution of the \Lya forest is a direct consequence of the 
universal expansion.}  The direct numerical results of Zhang et al. (1998)
support this conclusion. If we include the evolution of the baryon 
overdensity, as shown in Figure \ref{fig:overdens} for the sCDM simulation, 
and shift the overdensity distribution until the peaks overlap, the $
(1+\delta_{b})^{1.7}$ dependence in Equation \ref{eq:taufull} for the 
optical depth distributions predicts additional scaling factors 
of $\approx 1.64$ ($0.77$) at $z=2$ ($z=4$), respectively, for $\tau$ that 
bring the distributions (shown in the bottom panel of Figure 
\ref{fig:scdmtau}) into close agreement. The remaining small differences, 
the slight broadening of the distribution and a reduction in its peak amplitude with decreasing redshift,
most probably reflect the fact that the shape of the baryon overdensity 
distribution is also evolving slowly with $z$.

\begin{figure}
\epsfxsize=3.5in
\centerline{\epsfbox{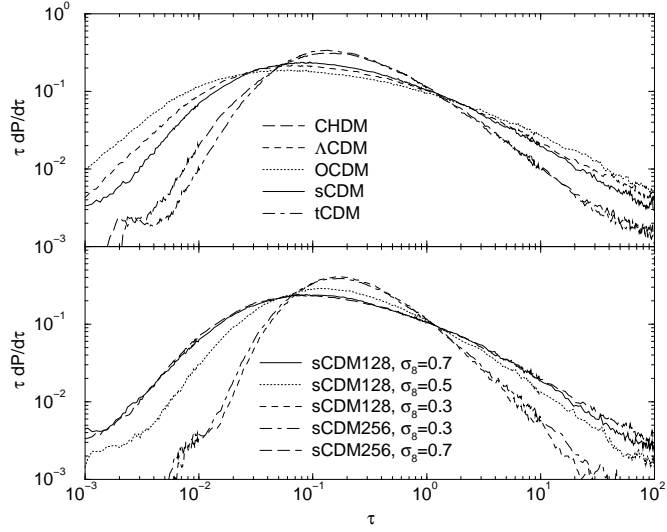}}
\caption{
Predictions for $\tau dP/d\tau$ for those cosmologies in the model comparison
study (top). Dependence of $\tau dP/d\tau$ on the fluctuation power spectrum 
in sCDM models with varying $\sigma_8$ and simulation grid resolution 
(bottom).}
\label{fig:tausplit}
\end{figure}

The top panel of Figure \ref{fig:tausplit}, shows $\tau dP/d\tau$ for the 
five sCDM simulations with varying power and spatial resolution.  From this 
we can see that the optical depth PDF at a given redshift is insensitive to 
the spatial resolution of the simulation. The shape of the distribution, 
however,  is strongly dependent on the amount of small scale power present.  
Models with less power at these scales produce narrow, sharply peaked 
distributions. As the power increases, the distribution flattens and 
broadens.  In the lower panel of Figure \ref{fig:tausplit} we show 
$\tau dP/d\tau$ versus $\tau$ for the simulations in the model comparison 
study. These distributions again display a clear dependence on the power 
spectrum of the model with OCDM (our model with the most small-scale power) 
producing the broadest distribution, and CHDM and tCDM (models with the 
least small-scale power) producing the most sharply peaked distributions. 
Thus this statistic is 
particularly promising as a model discriminator in that these differences 
between models are significant in the range $0.02 < \tau < 4$ that should be 
accessible to observers. 

\begin{figure}
\epsfxsize=3.5in
\centerline{\epsfbox{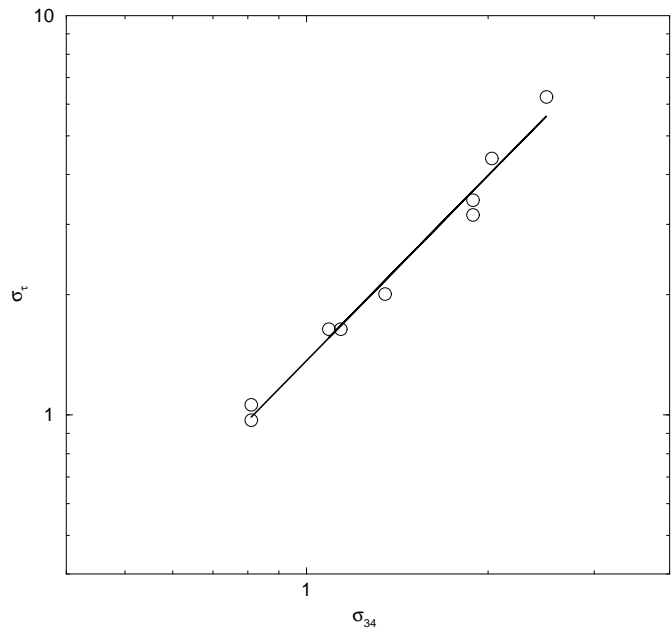}}
\caption{
Correlation of $\sigma_\tau$, a measure of the width of the $\tau dP/d\tau$ 
distribution, with $\sigma_{34}$, the amplitude of the linear power spectrum 
on small scales as given in Equation \protect\ref{eq:sig34}.}
\label{fig:tausig34}
\end{figure}

We quantify this relation between the shape of the $\tau$ distribution and 
the amplitude of the power spectrum by fitting a log-normal to the curves:
\begin{equation}
\tau {dP \over d\tau} \propto  e^{-(\ln\tau-\ln\tau_0)^2 / 2\sigma_\tau^2 }.
\label{eq:tauprox}
\end{equation}  
Although this does not fit the profiles in Figure \ref{fig:tausplit} in 
detail, it does provide an adequate description as long as we restrict the 
range of optical depths used in the fitting.  Here we adopt $0.02<\tau<4$,
corresponding roughly to the observable range.  A different range or a 
different fitting function changes the details, but not the nature of our 
result.  In Figure \ref{fig:tausig34}, we show the correlation between
$\sigma_\tau$, a measure of the width of the distribution, and $\sigma_{34}$,
the amplitude of the linear power spectrum on small scales as defined in 
Equation \ref{eq:sig34}. The strength of the correlation is striking.  The 
low scatter around the power law relation shown in this figure bolsters our 
claim that the shape of the $\tau$ distribution function is insensitive to 
other cosmological parameters.  To give an idea of the uncertainty in each 
point, we fit both the high and low resolution simulations for the sCDM 
$\sigma_8=0.3$ and $\sigma_8=0.7$ models.  In both of these cases 
$\sigma_\tau$ differs by less than 10 \%.
    
\subsection{Flux Probability Distribution}
\label{sec:flux}

\begin{figure}
\epsfxsize=3.5in
\centerline{\epsfbox{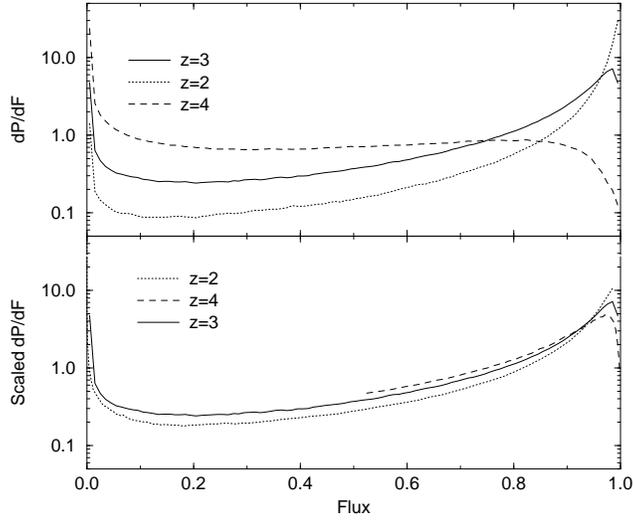}}
\caption{
(top)Flux probability distribution for the high spatial resolution sCDM 
model with $\sigma_{8}=0.7$ for $z=2,3$ and $4$. (bottom) The same sCDM
distributions as above where both the distributions and the flux are scaled 
to the $z=3$ values predicted by the simple scaling of $\tau$ given by 
Equation \protect\ref{eq:tausim}.} 
\label{fig:scdmfpdf}
\end{figure}

  Although the optical depth PDF is easier to model theoretically, the flux 
PDF (where $dP/dF$ is the probability that a pixel will have transmitted 
flux between $F$ and $F+dF$) is closer to actual observation.  The top panel 
of Figure 
\ref{fig:scdmfpdf} shows the flux probability distribution functions for 
the high spatial resolution sCDM model with $\sigma_{8}=0.7$ at $z = 2,3$ 
and $4$. The bottom panel of Figure \ref{fig:scdmfpdf} shows the prediction
of the simple $\tau$ scaling given in Equation \ref{eq:tausim} applied to 
the flux and these same flux probability distributions.  Again we attempt to 
rescale the $z=4$ and $z=2$ distributions to $z=3$ in order to test the 
scaling. This results in a highly nonlinear 
mapping of the flux and the flux PDF from $z=j$ to $z=3$ given by 
$dP_{j}/dF \rightarrow \eta F^{1-1/\eta}dP_{j}/dF$ and $F_{j} \rightarrow F_{j}^{1/\eta}$ for $\tau_{j} \rightarrow \tau_{j}/\eta$, where $\eta = 0.356(6.511)$ for $j=2(4)$, respectively. While the shapes of the 
distributions in the top panel appear quite different, much of the $z$ 
evolution of the flux probability distributions is explained by this simple 
scaling, the remainder representing mostly the effect of the evolution of the 
baryon density in the cosmological model.  We do not plot the scaled 
distribution for $z=4$ below the scaled flux of $0.5$ because this already 
corresponds to an unscaled flux of $0.015$, close to saturation and most 
likely noise dominated in the observations.

\begin{figure}
\epsfxsize=3.5in
\centerline{\epsfbox{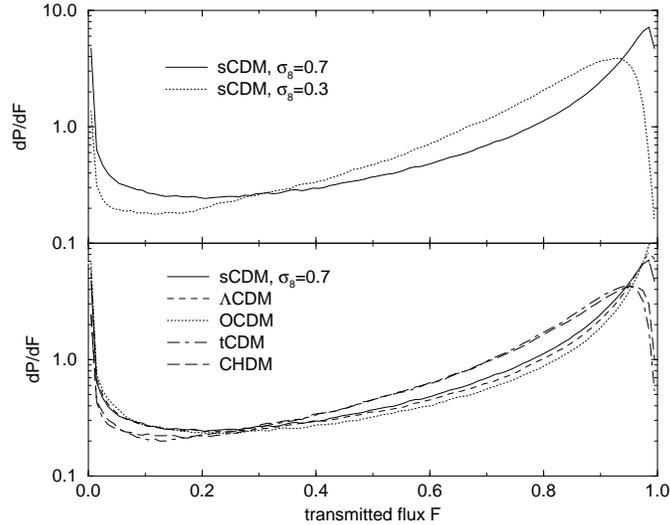}}
\caption{
(top)Flux probability distributions at $z=3$ for sCDM with spatial resolution 
$\Delta x=37.5$~kpc and $\sigma_{8}= 0.7$ and $0.3$, respectively. (bottom) 
Flux probability distributions at $z=3$ for the models in the model 
comparison study.}
\label{fig:fpdfmodels}
\end{figure}

The flux PDF depends only weakly on simulation grid resolution (Bryan 
\etal 1998). Its shape is strongly dependent on the power spectrum of the 
underlying cosmology. In the top panel of Figure \ref{fig:fpdfmodels}
we show the flux probability distributions in the sCDM model (spatial 
resolution $\Delta x = 37.5$~kpc) for cluster scale normalizations 
$\sigma_{8}=0.3$ and $0.7$. The dependence on the normalization of the 
power spectrum is clear.  The number of pixels 
found with flux in the central flux range $0.3 < F < 0.9$ is greater for 
models with less power ($\sigma_{8}=0.3$); while the number of pixels with 
flux in the low ($F< 0.3$) and high 
($F>0.9$) flux ends of the distribution are less than for models with 
greater power ($\sigma_8=0.7$).  This 
is in qualitative agreement with Croft \etal (1997a).  We note, however, that 
our result (using the Kronos code) does not require any smoothing of the 
simulations as was the case for their TreeSPH simulations. In the lower 
panel of Figure \ref{fig:fpdfmodels} we present the $z=3$ flux 
PDFs for the five models of the model comparison study. Models with lower 
power at small scales (tCDM, CHDM) have a larger flux PDF for $0.3<F<0.9$ 
than sCDM and \LCDM, while the low density model (OCDM) with the highest 
spectral power at these scales has the smallest flux PDF in this range, as 
expected. Furthermore, the differences between models can be substantial.
For example, at $F=0.6$ the OCDM results lie $10$ \% below the \LCDM model 
result while the CHDM result lies above the \LCDM result by about a factor 
of $1.4$. We remind the reader that the mean of the distribution has been 
fixed to match observations. Thus this statistic should be useful to 
constrain competing models. 

\subsection{Fraction of high \Lya opacity}
\label{sec:fraction}

Another possibly useful statistic for discriminating models is the fraction
of a quasar spectrum with \Lya optical depth greater than a specified
value $\tau_0$, i.e. the cumulative distribution in optical depth.
Small differences in the amplitude normalization of the primordial
power spectrum may be enhanced in the cumulative opacity data (Cen 1997).

\begin{figure}
\epsfxsize=3.5in
\centerline{\epsfbox{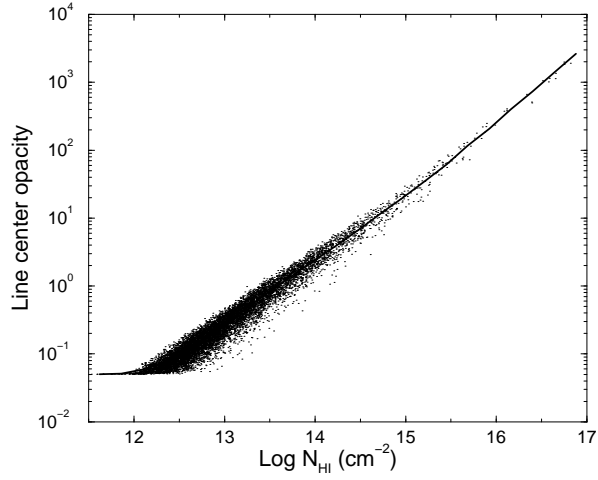}}
\caption{
Scatter plot of the line center opacity as a function
of column density for the sCDM model run with the Hercules
code. Also shown (solid line) is the average line center
opacity in the 24 density bins of width
$\Delta \log N_{HI} = 0.24$ \cms.
The cutoff in opacity at $\tau_c > 0.05$ is due to the
spectral threshold $e^{-\tau} = 0.95$ used in identifying
the absorption features.}
\label{fig:tauvsNH}
\end{figure}

\begin{figure}
\epsfxsize=3.5in
\centerline{\epsfbox{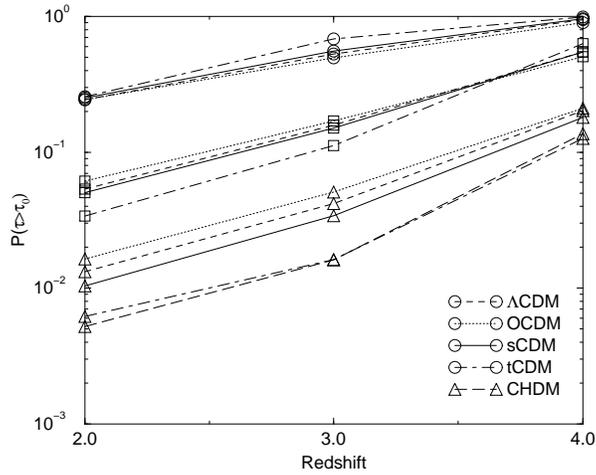}}
\caption{
Fraction of pixels greater than three threshold optical depths,
$\tau_0=0.1$ (circles), $\tau_0=1.0$ (squares), and $\tau_0=7.0$
(triangles), plotted against redshift. The curves for tCDM and CHDM are 
indistiguishable for the lowest two thresholds. }
\label{fig:tauthresh}
\end{figure}

\begin{figure}
\epsfxsize=3.5in
\centerline{\epsfbox{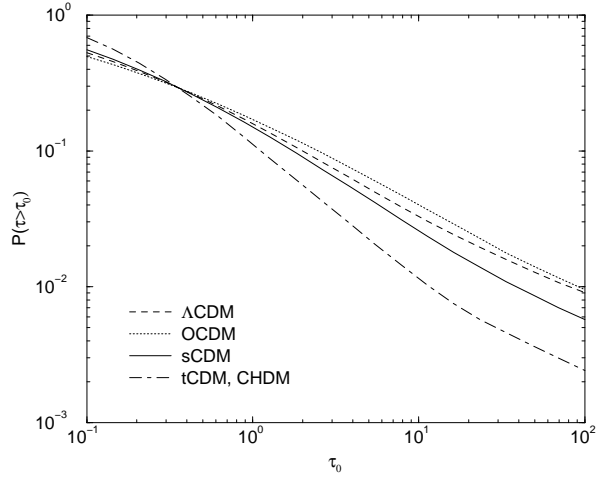}}
\caption{
Fraction of pixels exceeding a threshold optical depth $\tau_0$ at $z=3$ 
for models in the model comparison study. The curves for tCDM and CHDM are 
indistinguishable.}
\label{fig:tauthz3}
\end{figure}

Figure \ref{fig:tauvsNH} shows the linear correlation between
the opacity at line center and the column density of absorption
features in the sCDM model, ranging from the optically thin
to thick at the Lyman edge. The nearly
unbroken relation $\tau_c \propto N_{HI}$, which
exists down to the incompleteness density of $\sim 10^{12}$ \cms
is attributed to the weak correlation between the Doppler parameter
and column density since, in general, $N_{HI} \propto b \tau_c$.
The lower bound on the opacity ($\tau_c > 0.05$) is set by the
transmission or spectral threshold $F_t = e^{-\tau_t} = 0.95$ used in the
line identification procedure.
Using Figure \ref{fig:tauvsNH} as a guide, we investigate the
cumulative opacity distribution with the following minimum opacity
thresholds: $\tau >$ 0.1, 1, and 7 which, if
associated with the line centers, would correspond roughly to
column densities of $\log N_{HI} =$ 12.5, 13.5 and 14.5 respectively.
The distributions $P(\tau>\tau_0)$ for the above minimum opacity thresholds 
are plotted in Figure \ref{fig:tauthresh}
at redshifts $z=2$, $3$, and $4$ for the models in the model comparison study.
In comparing groups with the same $\tau_0$, the smaller threshold curves
are more highly clustered and less sensitive to the background cosmological
model parameters. This is especially evident in Figure \ref{fig:tauthz3}
where we show the cumulative distributions of the optical depth at redshift
$z=3$ for these models. 


\section{Line Parameter Statistics}
\label{sec:spectrum}

In this section we present a line analysis of the spectra generated from 
the various model simulations. 
We compare and contrast the cosmological models based on the column density
distribution and the evolution of line number.

\subsection{\HI Column Density Distribution}
\label{sec:column}

\begin{figure}
\epsfxsize=3.5in
\centerline{\epsfbox{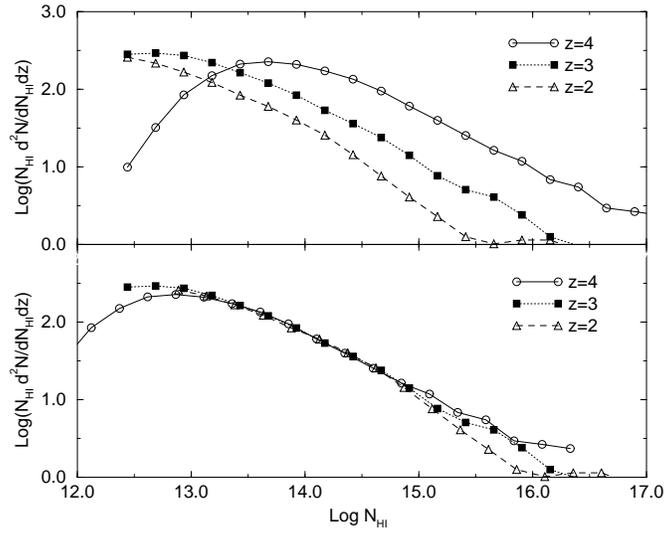}}
\caption{
(top) Uncut \HI column density distributions for the high resolution sCDM 
$\sigma_{8}=0.7$ model for redshifts $z=2,3$ and $4$; (bottom) The same 
\HI column density distribtions with $\NHI$ scaled according to 
Equation \protect\ref{eq:NHIscale} to $z=3$.} 
\label{fig:raw_nlgtNH}
\end{figure}

\begin{figure}
\epsfxsize=3.5in
\centerline{\epsfbox{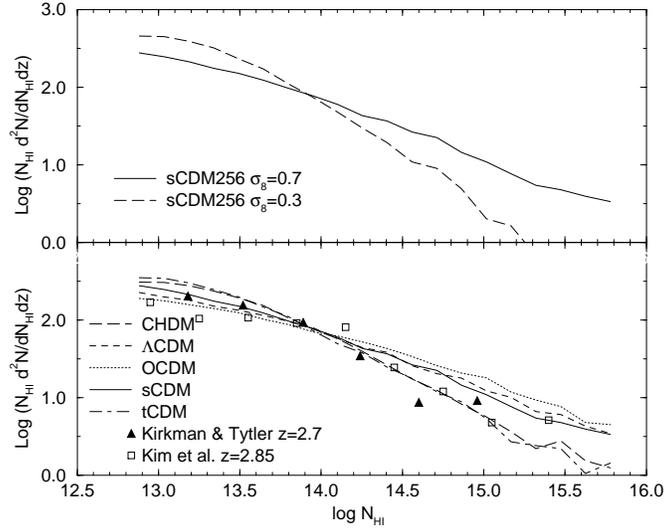}}
\caption{
(top) \HI column density distribution for high spatial resolution 
($\Delta x=37.5$~kpc) sCDM models with $\sigma_{8}=0.3$ (dashed) and 
$\sigma_{8}=0.7$ (solid). (bottom) \HI column density distributions
at redshift $z=3$ for (Kronos) simulated spectra in the model comparison 
study. Observational data are from Kirkman \& Tytler (1997) and Kim \etal 
(1997).}
\label{fig:nlgtNH}
\end{figure}

One of the most robust line statistics used in the analysis of the 
\Lya forest is the \HI column density distribution, which is well converged 
by simulation box sizes of $9.6$ Mpc and is insensitive to changes in the 
simulation grid resolution or treatment of gas hydrodynamics (Bryan \etal 
1998, Zhang \etal 1997).  The \HI column density distribution, defined
to be $N_{HI}=\int_{x_A}^{x_B}{n_{HI}\over 1+z} dx$,  is closely 
related to the optical depth $\tau$ through the dependence of each on the 
number density of neutral hydrogen.  Thus using Equation \ref{eq:nHI} and 
the approximations that led to Equation \ref{eq:taufull} we 
expect the \HI column density to scale as
\begin{equation}
N_{HI} \propto {(\Omega_B h^2)^2 \over \Gamma(z)H} (1+z)^{4.8}\int
(1+\delta_{b})^{1.7}\,du
\end{equation} 
In the top panel of Figure \ref{fig:raw_nlgtNH} we show the raw (uncut) 
\HI column density distribution for the high resolution sCDM model for 
redshifts $z=2,3$ and $4$. In the bottom panel of Figure \ref{fig:raw_nlgtNH}
 we see that the column density scaling 
\begin{equation}
N_{HI} \propto {(\Omega_B h^2)^2 \over \Gamma(z)H} (1+z)^{4.8},
\label{eq:NHIscale}
\end{equation}
the same relation as the naive scaling relation given in Equation 
\ref{eq:tausim} for the optical depth $\tau$, accounts for the redshift 
evolution of the column density distribution amazingly well.  This 
demonstrates that the column density, an integrated quantity, is much less 
sensitive than the optical depth distributions to the redshift evolution of 
the gas overdensity within an absorbing structure.  The differences 
seen in the low column density end of the distributions, particularly for 
$z=4$, may be a result of the simulation spatial resolution 
(Bryan \etal 1998), while the differences observed in the high column density
end may partially be due to shot noise in the high $z$ data.  
In Figure \ref{fig:nlgtNH} (top) we 
explore the dependence of this distribution at a given redshift ($z=3$) on 
the power spectrum of the underlying cosmology. We find qualitative 
agreement with semi-analytic arguments (Gnedin 1998; Hui, Gnedin, \& Zhang 
1997) in that models with less power on small scales (such as  sCDM with 
$\sigma_8=0.3$ and 
$\sigma_{34}=0.812$) have \HI column distributions with significantly 
steeper slopes than models (such as sCDM with $\sigma_8 = 0.7$ and 
$\sigma_{34} = 1.89$) with more power at these scales.  However, as we 
discuss in more detail below and in Table \ref{tab:beta}, quantitative 
agreement between the simulations and the predictions of these  
semi-analytic arguments seems more difficult to achieve.  

In Figure \ref{fig:nlgtNH} (bottom), we show the \HI column 
density distribution at redshift $z=3$ for (Kronos) simulated spectra in the 
model comparison study and compare the simulated data with data from Kirkman
\& Tytler (1997) 
and the fits provided by Kim \etal (1997).
The distributions are conventionally quantified by fitting them to power
laws, $dN/dN_{\rm HI} \propto \NHI^{-\beta}$. We use the same sets of column 
density cuts on the simulated data in the model comparison study as Kim \etal
in order to expedite comparison with the data and use a direct unweighted 
least squares fit (all quoted errors are $2\sigma$) to extract the slope 
$\beta$ from the simulated data.  
Our results are summarized in Table \ref{tab:beta}.
We find again the expected dependence on the fluctuation power spectrum.  For 
the column density range $13.7<\log\NHI<14.3$ (given by the column labeled 
$\beta_{\rm h}$) the most shallow slope is for 
OCDM, the low matter density model with $\sigma_{34}=2.50$  while CHDM and 
tCDM with $\sigma_{34}= 1.14$ and $1.09$, respectively, give the steepest
distributions (see Table \ref{tab:sim_par}). The predicted column density 
distributions generally also steepen with time (decreasing redshift). 
Kim \etal find $\beta=1.46\pm0.07$ ($2\sigma$) for
this column density range at $\langle z\rangle=2.85$. This is formally
inconsistent with all the models at the $3\sigma$ level except for OCDM,
although it is marginally consistent with \LCDM.

Results for the column density range $12.8<\log\NHI<14.3$ in Table \ref{tab:beta} are shown in the column labeled $\beta_\ell$. The average distributions 
are generally shallower when extended to lower column densities, showing that the distributions are
curved. Kim \etal similarly find a shallower distribution over this column
density range with their results for lines at $\langle z\rangle=2.31$, 
$\langle z\rangle=2.85$, and $\langle z\rangle=3.35$ shown in the last row 
of Table \ref{tab:beta}.  The result at $\langle z\rangle=2.31$ is 
inconsistent with all of the simulation results at $z=2$, but note that the 
quoted uncertainty in the observation is eight times smaller than at 
$\langle z\rangle=2.85$, despite comparable numbers of 
absorbers.  By contrast, the result at $\langle z\rangle=2.85$ is formally
consistent at the $3\sigma$ level with all the models. At $\langle 
z\rangle=3.35$, Kim \etal find $\beta=1.59\pm0.13$ ($2\sigma$). This result 
is consistent with the simulation results for the OCDM model and marginally 
consistent (at the 
$3\sigma$ level) for the sCDM and \LCDM models.  The observational data, 
however, also suggest a weak steepening of the distribution with increasing
redshift, contrary to our findings. These discrepancies might indicate that
the redshift evolution of the ionizing radiation field may be somewhat 
different from that of the Haardt \& Madau spectrum assumed in the 
simulations.
 
We may compare the simulation
results with the semi-analytic predictions of Hui, Gnedin, \& Zhang
(1997) to understand the trend of changing steepness with power
spectrum. We provide the predicted values of $\beta$ according to the
prescription of Hui \etal in Table \ref{tab:beta}.  We assume
$T\propto\rho_B^{0.5}$, as found by Zhang \etal (1998) for this column
density range in an sCDM simulation. The uncertainty in the $T-\rho_B$
relation introduces only a 10\% uncertainty in the prediction for
$\beta-1$, so it seems reasonable to retain it for the other models as well
for this purpose. The predicted values of $\beta$ for the sCDM, tCDM, and
CHDM models at $z=3$ match the simulation values to within $1\sigma$, in 
agreement with the comparison in Hui \etal with one of our earlier sCDM 
models. However, the predictions for OCDM and \LCDM at $z=3$ are in 
disagreement with the semi-analytic arguments giving too steep a slope.  The 
predictions do less well for all models at $z=2$. In particular, the 
simulation results show a steepening of the column density distribution 
toward decreasing redshifts for all the models, opposite to the predicted
trend.  

Over the wider column density range $10^{12.8}<\NHI<10^{16}$ (summarized in 
the column labeled $\beta_{\rm f}$ in Table \ref{tab:beta}), we see that 
the average distributions continue to steepen toward higher column densities. 
Kim \etal obtain $\beta=1.46$ for this column density
range at $\langle z\rangle=2.85$, with a steepening to $\beta=1.55$ at
$\langle z\rangle=3.7$. The results for the tCDM and CHDM models 
($1.95\pm0.06$ and $1.92\pm0.06$ at $z=3$, respectively) are
substantially steeper than these values.  Because the distribution deviates 
from a pure power law at the low column density end, it is useful to split 
the simulation samples into two halves, fitting each to a power law. These 
results are given in the last two columns of Table \ref{tab:beta} where 
$\beta_1$ is the slope of the column density distribution for lines at the 
low column density end ($10^{12.8} < \NHI < 10^{14}$) and $\beta_2$ is the 
slope of the column density distribution for lines at the high end 
($10^{14}<\NHI<10^{16}$).  Giallongo \etal (1996) obtain $\beta=1.8$ for 
systems with $\NHI> 10^{14}$ and $2.8<z<4.1$. 

Finally we note that the analogous column density distributions derived from 
the lower resolution Hercules runs give similar results
and slopes as the Kronos data. For example over the full column density range
at $z=3$, Hercules data give slopes for the column density distribution of 
$1.71$, $1.66$ and $1.62$ for the sCDM, \LCDM and OCDM models, respectively, 
consistent within errors with the Kronos results.  This suggests that the
distribution function is a robust diagnostic, 
being relatively insensitive to grid resolution and numerical method.  
A preliminary comparison with the data favors models with more power at these
scales than in our CHDM or tCDM cosmologies.  However, there appears to be 
some discordance in the observations, so a more definitive comparison will 
require more work.

\subsection{Line Number Evolution}
\label{sec:linenumber}

The number of \Lya lines at a particular redshift reveals 
how many intergalactic absorbers exist at that time between
the quasar and observer and, given certain assumptions on their
geometry, the size and volume filling factor of the absorbers
can also be deduced. Since the column density of \Lya lines
corresponds to the mean overdensity and size of the clouds fairly well
(Charlton \etal 1997; Zhang \etal 1997), it is useful
to see how the number of lines evolves with different column density cutoffs,
as this will track the evolution of morphologically distinct
small scale structures in the universe.

\begin{figure}
\epsfxsize=3.5in
\centerline{\epsfbox{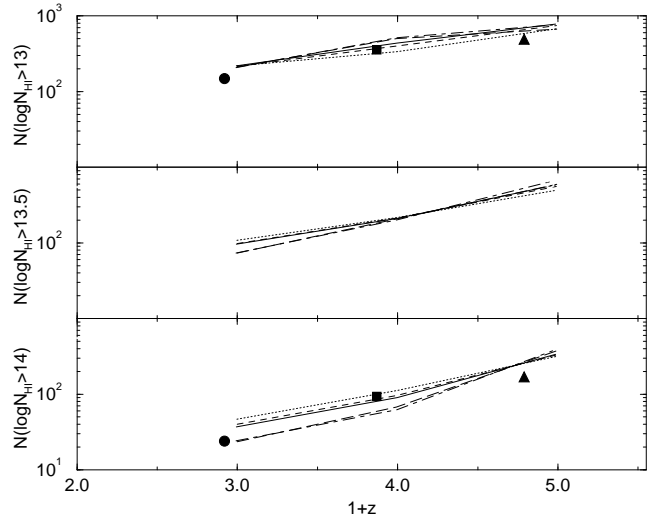}}  
\caption{
Evolution of the number of lines with 
column densities greater than $\log \NHI >13$(top), $\log \NHI >13.5$
(middle), $\log \NHI >14$ (bottom) per unit redshift for  the 
sCDM $\sigma_8=0.7$ (solid), \LCDM (dashed), OCDM (dotted), 
tCDM (dot-dashed), and CHDM (long dashed) models of the model comparison 
study. The numerical results are compared against the observed data of
Kulkarni \etal 1996 (filled circle), Hu \etal 1995 (filled square) and 
Lu \etal 1997 (filled triangle).}
\label{fig:mod_nl13z}
\end{figure}

Figure \ref{fig:mod_nl13z} shows the evolution of the number of lines with 
\HI column densities greater than $10^{13}$ \cms,
$10^{13.5}$ \cms , and $10^{14}$ \cms , respectively, comparing
results for the models in the model comparison study with the
observed data from Kulkarni \etal (1996) at $z\sim 2$,
Hu \etal (1995) at $z\sim 3$ and Lu \etal (1997) at $z\sim 4$.
For a fixed transmission cutoff
(here $F_t=0.95$) and column density threshold,
the total number of lines per unit redshift decreases
with time because the opacity of the universe
decreases from both the increasing flux of radiation and the expansion
of the universe.  
With the exceptions of the tCDM and CHDM models for the lowest column density 
threshold $\NHI > 10^{13}$ where incompleteness due to line blending becomes 
significant at higher $z$, the deviation from a fixed power law behavior 
tracks predominantly the behavior of the radiation flux. 
Fitting the evolutions to the form $dN/dz \propto 
(1+z)^\gamma$ over the range $2<z<4$, we find the exponents are fairly 
similar in the different models.  We summarize these results in Table \ref{tab:gamma} (all errors are $2\sigma$). To compare these simulated results with
observational data we fit the combined line lists from Kulkarni \etal, 
Hu \etal, Kirkman \& Tytler, and Lu \etal
to the same power law behavior and display those results in the row labeled 
``combined'' in Table \ref{tab:gamma} again with $2\sigma$ errors. 
(Lines near the QSO emission redshift were avoided because
of the proximity effect, as were lines associated with metal systems.) 
Kim \etal obtain $\gamma=2.78\pm1.42$ ($2\sigma$),
fit over $2<z<3.5$ for systems with column densities
$10^{13.77}<N_{\rm HI}<10^{16}$ \cms. Using the same column density 
cuts and simulation data for $z=2$ and $3$ only, we find power law exponents 
(labeled $\gamma_{\rm h}$ in Table \ref{tab:gamma})
for the sCDM, \LCDM, OCDM, tCDM and CHDM models of the comparison study in 
good agreement with the observational result. For the lower column density 
range $10^{13.1}<N_{\rm HI}<10^{14}$ \cms, the power law exponents are labeled
$\gamma_\ell$ in Table \ref{tab:gamma}. Kim \etal (1997) obtain 
$\gamma=1.29\pm0.90$ ($2\sigma$) fit over $2<z<4$. Our sCDM, \LCDM and OCDM 
model predictions are in good agreement with this observational result, 
although tCDM and CHDM show a somewhat stronger evolution.

\begin{figure}
\epsfxsize=3.5in
\centerline{\epsfbox{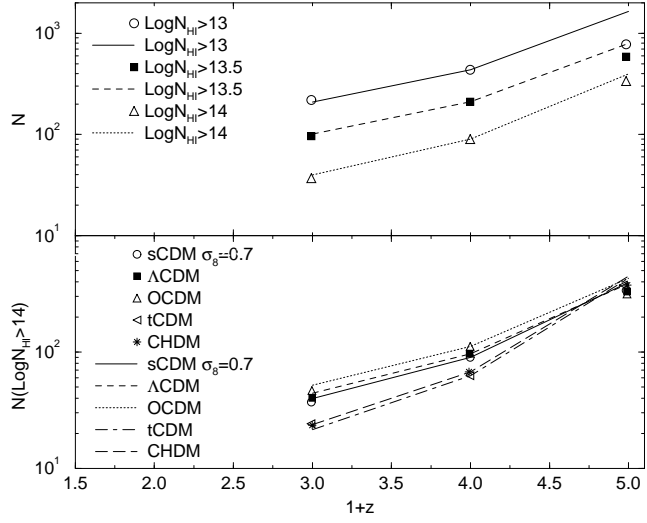}}
\caption{
(top) Scaling predictions (lines) for the line number evolution with 
redshift for lines with column densities greater than $\log \NHI >13$, 
$\log \NHI >13.5$, $\log \NHI >14$ per unit redshift computed for the high 
spatial resolution sCDM $\sigma_8=0.7$ model and compared to the simulation 
data (symbols).  The scaling prediction is normalized to the simulation 
data at $z=3$. (bottom) Scaling predictions (lines) for the line number 
evolution with redshift for lines with column densities greater than 
$\log \NHI >14$ for the models in the model comparison study compared
to simulation data (symbols) for  the same models.}
\label{fig:dndzscale}
\end{figure}

All the models yield comparable levels of evolution
at each of the  column density cutoffs and, in fact, the evolution slopes in
all the models agree for the most part within errors with the observed 
values. Two trends in the model predictions are  apparent.  
The first is that at a given column density threshold the slope $\gamma$ of 
the line number evolution is correlated to the slope $\beta$ of the column 
density distribution with $\gamma$ increasing for models with larger $\beta$ 
(i.e. models with less power in the fluctuation spectrum at small scales). 
The second trend is that stronger column density lines for a given model 
exhibit a greater rate of redshift evolution. This is not unexpected since 
the evolution for a fixed transmission threshold is essentially determined by 
the radiation field (which is the same in all the models) and, 
to a lesser degree, the expansion of the universe.  In our previous studies 
(Zhang \etal 1997) we have found little intrinsic cloud evolution over these 
redshift intervals.  If we assume that the evolution of the gas overdensity 
does not contribute significantly to the evolution of line number or column 
density at these redshifts we can use Equation \ref{eq:NHIscale} and the 
power law dependence of the column density distribution 
$dN/(dN_{\rm HI}dz) \propto \NHI^{-\beta}$ to predict the number of lines 
above a fixed column density threshold as a function of redshift $z$.  We 
find  
\begin {equation}
{dn \over dz}(>N_{HI}) \propto \Bigg( {(1+z)^{4.8} \over \Gamma(z)H(z)} \Bigg)^{(\beta -1)}
\label{eq:dndzscale}
\end {equation}
In Figure \ref{fig:dndzscale} we compare the scaling predictions of Equation 
\ref{eq:dndzscale} to the simulation results using the high resolution sCDM 
$\sigma_8 = 0.7$ model for column density thresholds $\NHI > 10^{13}$, 
$10^{13.5}$, and $10^{14}$ \cms, respectively.  Since $\beta$ also evolves 
weakly with $z$, we use $\beta$ at $z=3$ as representative of the average for
$2<z<4$ in Equation \ref{eq:dndzscale}.  For the lowest two column density 
thresholds we use the single power law fit to the column density distribution
over the range $10^{12.8}<\NHI<10^{16}$ \cms, while we use $\beta_2$ from the
 two 
power law fit for the high ($\NHI >10^{14}$ \cms) column density threshold.  
We normalize the scaling predictions to the simulated number of lines at 
$z=3$ because that is where all the models in our study were normalized to 
the observational data.  For the lower two column density 
thresholds, the scaling prediction tends to overestimate the number of lines 
at $z=4$.  For the lowest column density threshold ($ \NHI > 10^{13}$) 
this again is partly due to incompleteness in the simulation line lists 
caused by line blending, an effect that becomes more severe for low column 
densities at high $z$.  Furthermore the slope of the low column density end 
softens 
for $\NHI <10^{14}$  with the break at $\NHI \sim 10^{14}$ probably 
reflecting a change in the absorbers from low density structures evolving 
primarily with the universal expansion to structures undergoing gravitational 
collapse (Bryan \etal 1998).  This deviation of the column density 
distribution from the pure power law assumed in the scaling relation would 
also cause the scaling law to overproduce the low column density lines. 
For the column density threshold  $\NHI > 10^{13.5}$ the discrepancy between 
the scaling prediction and the simulation results at $z=4$ is reduced. This  
is to be expected since the high spatial resolution sCDM linelists should be 
complete at this column density threshold. For lines with $\NHI > 10^{14}$, 
where the absorbers share a common morphological type and the 
column density distribution is well fit by a single power law, agreement 
between the scaling prediction and the simulations is good.  In the lower
panel of Figure \ref{fig:dndzscale} we compare the scaling predictions with 
the simulation results for the models in the model comparison study for the 
high column density threshold case.  The scaling predictions for all models
agree reasonably well with the simulations.

\begin{figure}
\epsfxsize=3.5in
\centerline{\epsfbox{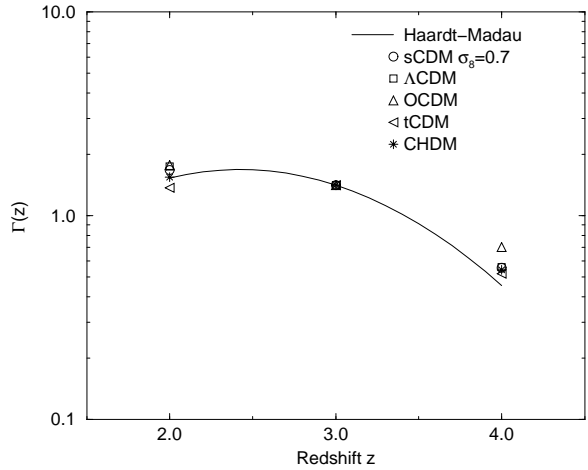}}
\caption{
Scaling law predictions of the redshift evolution of the metagalactic UV 
radiation field  for models in the model comparison study compared to that 
of the Haardt--Madau field used in the simulations.  The amplitude of the 
predicted radiation field is normalized to agree with the Haardt--Madau 
field at $z=3$.}
\label{fig:gammapred}
\end{figure}

Encouraged by these results we solve Equation \ref{eq:dndzscale} for the
shape of the UV ionizing background $\Gamma (z)$ in terms of the Hubble 
parameter $H(z)$ modeling the universal expansion and the (in principle)
measurable quantities $dn/dz(\NHI >10^{14})$ and $\beta_2$, the slope of the 
column density distribution over this column density range.
\begin{equation}  
\Gamma (z) \propto {(1+z)^{4.8} \over H(z)}\left( {dn  \over dz} \right)^{1/(1-\beta_2)}
\label{eq:gammaz}
\end{equation}  
We use Equation \ref{eq:gammaz} to compute $\Gamma (z)$ with simulation data 
from the model comparison study and compare these predictions to the 
Haardt--Madau spectrum actually used in Figure \ref{fig:gammapred}. Although 
the prediction is highly sensitive to the slope of the column density 
distribution used (whose errors are still quite large), it is gratifying, 
given the simplicity of the scaling relations, that all of the models 
reproduce the assumed Haardt-Madau evolutionary trend.


\section{Doppler $b$ Parameter}
\label{sec:bparm}

   Recent papers (Bryan \etal 1998, Theuns \etal 1998) have shown that both 
the Doppler $b$  parameter and a related nonparametric statistic, the mean 
flux difference as a function of velocity, require high simulation spatial 
resolution to model properly. In this section we investigate the 
dependencies of these statistics on the properties of the cosmological model.
 Because these statistics are highly sensitive to the spatial resolution of 
the simulation, we present results only for those models (sCDM, \LCDM, OCDM, 
and tCDM)  simulated with our highest spatial resolution $\Delta x=37.5$~kpc. 

\begin{figure}
\epsfxsize=3.5in
\centerline{\epsfbox{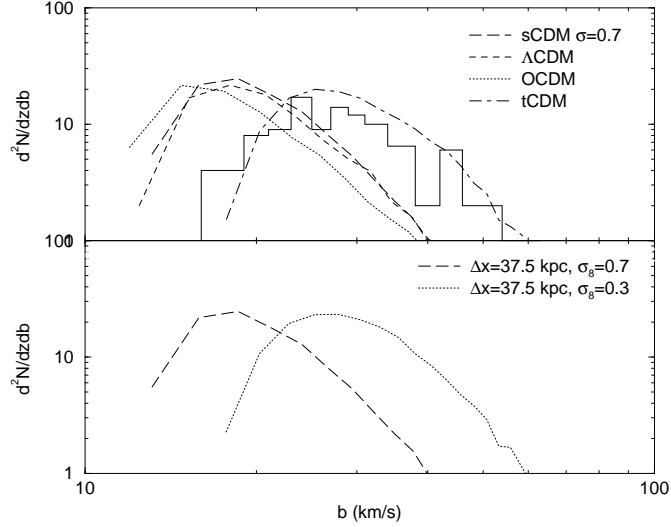}}
\caption{
Doppler parameter distributions at redshift $z=3$ from
the high resolution ($\Delta x = 37.5$~kpc) simulated data. 
(top) Doppler parameter distributions for sCDM, \LCDM, OCDM, tCDM 
models of the model comparison study. The histogram is data from 
Kim \etal (1997) for $z=3.35$. (bottom) Doppler parameter 
distributions for sCDM $\sigma_8=0.3$ and $0.7$ to investigate the 
dependence on the fluctuation power spectrum.} 
\label{fig:mod_nlb}
\end{figure}

The Doppler $b$ parameter measures the amount of line broadening due to 
thermal broadening, physical velocities, Hubble expansion broadening and the 
shape of the absorber density profile (Bryan \etal 1998).  Both Hubble and 
thermal broadening are significant for the lower column density lines that
arise from structures found in voids that are still expanding in 
absolute coordinates.  The thermal contribution only becomes dominant for the 
higher column density lines that have turned around and are gravitationally 
collapsing. Furthermore, the $b$ parameter is highly sensitive to the 
simulation spatial resolution.  Lower resolution simulations numerically 
thicken the lines causing the width of the lines, and thus $b$ to be 
overestimated (Bryan \etal 1998; Theuns \etal 1998). In our previous work 
(Bryan \etal 1998) we argued that the shape of the $b$--distribution was in 
rough agreement with observation and particularly that the high $b$ power 
law tail of the distributions arises naturally in hierarchical models when 
quasar lines of sight pass obliquely through the filamentary absorbing 
structures (Rutledge 1998).  However the median of the simulated 
$b$ parameter distribution for the sCDM model, calculated from simulations 
with high spatial resolution, was now substantially below the 
$\sim 30$~km/s median seen in the observations. Thus   
the sCDM model, which previously had appeared to be in agreement with 
observations, is now discrepant.  In  Figure \ref{fig:mod_nlb} we show the 
Doppler $b$ parameter distributions extracted from the high grid resolution 
($\Delta x= 37.5$~kpc) Kronos simulations at redshift $z=3$ for the sCDM, 
\LCDM, tCDM, and OCDM models for lines with column densities between 
$10^{13.1}$ \cms $< \NHI < 10^{14}$ \cms. We present for comparison data 
from Kim \etal (1997) for $z=3.35$.  The \LCDM and OCDM models, like 
sCDM, have their $b$ distributions shifted too much to the left 
(to low $b$ values) to agree with observation.  Only tCDM, the model with 
the least fluctuation power at small scales and thus broader density 
structures at this redshift, has a 
median $b$ approaching the observational values.  We explore this dependence 
on the fluctuation power spectrum with the two highest resolution sCDM models 
(with $\sigma_8=0.7$ and $0.3$, respectively) in the lower panel of 
Figure \ref{fig:mod_nlb} and see that indeed the model with lower spectral 
power produces a $b$ parameter distribution shifted towards higher $b$ 
(as predicted by Hui \& Rutledge 1998).  The increase in $b$ for 
models with less fluctuation power at small scales may be partly due to line 
blending effects at these low column densities. However as shown below, the 
shift to higher $b$ values for these models persists for lines 
with higher column densities as well where line blending should not be as 
significant and thus can not be explained by line blending alone.    

\begin{figure}
\epsfxsize=3.5in
\centerline{\epsfbox{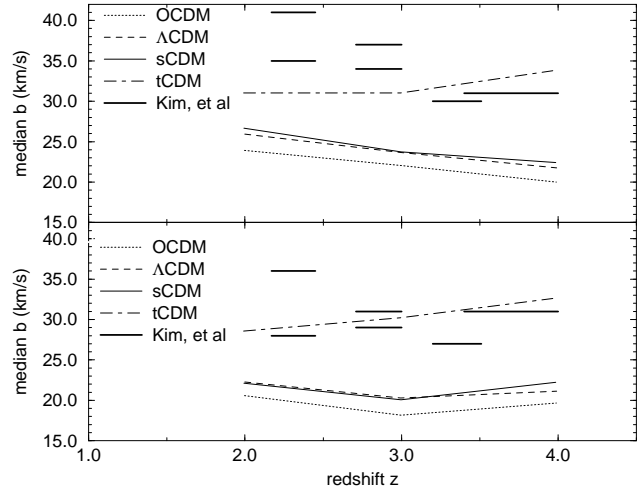}}
\caption{
Evolution of the median Doppler parameters for lines with column densities 
$10^{13.8}<\NHI<10^{16}$ (top) and $10^{13.1}<\NHI<10^{14}$ (bottom) for 
the high resolution models of the model comparison study.  Thick solid 
lines represent data from Kim \etal (1997).}
\label{fig:bmed_z}
\end{figure}

To facilitate a better comparison of the models with observations we plot the
median Doppler parameters as a function of redshift in 
Figure \ref{fig:bmed_z} where we have imposed the same column density
cuts on the lines as those used by Kim \etal (1997). The 
median $b$ for lines with column densities $10^{13.8}$ \cms $<\NHI<10^{16}$ 
\cms and 
$10^{13.1}$ \cms $<\NHI<10^{14}$ \cms are shown in the top and bottom 
panels, respectively.
While the \LCDM, sCDM, and OCDM models predict roughly the observed 
evolutionary 
trend for both sets of column density cuts, the median $b$ values lie 
systematically more than $6$~km/s below the observational data. OCDM, the 
model with the most power at these scales, is the most discrepant.   
Although tCDM predicts median $b$ parameters more consistent with observation,
the redshift evolution predicted by this model appears to be in disagreement 
with the data.  We can compare these results with other recent data sets.    
Confining to lines with $\NHI>10^{13}$ \cms, we obtain from the published 
line lists, for
$1.9<z<2.0$ ($1\sigma$ errors),
$(b_{\rm mean}, b_{\rm median})=(32.1\pm2.6, 29.7\pm3.3)$ \kms
($1\sigma$) (Kulkarni \etal), $3<z<3.1$ $(b_{\rm mean}, b_{\rm median})=
(38.0\pm1.6, 33.6\pm2.0)$ \kms (Hu \etal), $(b_{\rm mean}, b_{\rm median})=
(27.3\pm1.9, 25.9\pm1.2)$ \kms (Kirkman \& Tytler), and at $4<z<4.1$
$(b_{\rm mean}, b_{\rm median})=(32.6\pm2.4, 25.9\pm3.1)$ \kms (Lu \etal), 
again clearly discrepant with the model predictions. Thus none of the models 
considered here can restore agreement with the observational data.

\begin{figure}
\epsfxsize=3.5in
\centerline{\epsfbox{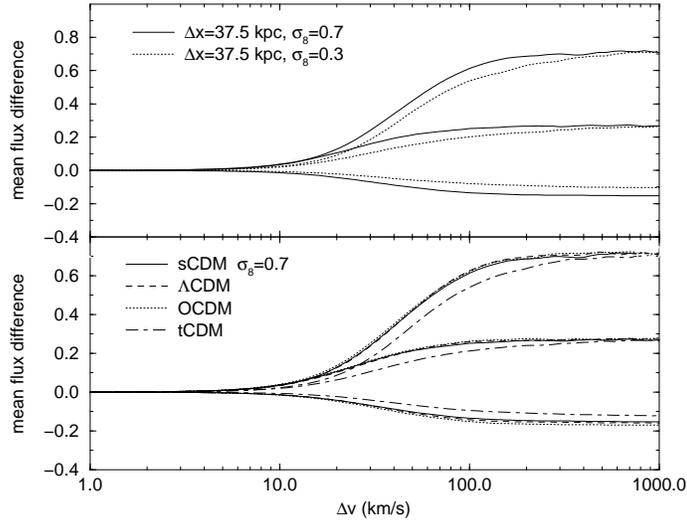}}
\caption{
The average flux difference as a function of velocity for pixels in the 
flux range (from top to bottom in each panel) $0.0<F<0.1$, $0.3<F<0.6$, and
$0.6<F<1.0$ for the high resolution sCDM $\sigma_8=0.7$ and 
$\sigma_8=0.3$ models (upper panel) and the high resolution 
models of the model comparison study (bottom panel).}
\label{fig:2point}
\end{figure}

We also argued in Bryan \etal (1998) that this discrepancy is not a result of 
the particular choice of line fitting algorithm, but appears for sCDM in the 
nonparametric moments of the two point flux distribution functions as well.
The two-point function $P_2(F_1,F_2,\Delta v)$ gives the probability that 
two pixels with separation $\Delta v$ will have flux $F_1$ and $F_2$. We 
plot the normalized moments of this function averaged over the flux range
$F_a$ to $F_b$  given by 
\begin{equation}
\label{eq:f2}
{
{\int_{F_a}^{F_b} dF_1 \int_0^1 dF_2 P_2(F_1,F_2,\Delta v)(F_1-F_2)}
\over
{\int_{F_a}^{F_b} dF_1 \int_0^1 dF_2 P_2(F_1,F_2,\Delta v)}
}
\end{equation}
which represents the average flux diffence as a function of velocity for 
pixels in the range $F_a$ to $F_b$.  
In Figure \ref{fig:2point} we plot the above statistic at $z=3$ as a 
function of velocity for several flux ranges for the 
high resolution models (sCDM, \LCDM, OCDM, tCDM) of the model comparison 
study (lower panel) and study its dependence on the small scale fluctuation 
power (top panel) using sCDM with $\sigma_8=0.7$ and $0.3$, respectively.   
There is little difference for low velocities, independent of flux level, due 
to the high coherence of the lines.  At very large velocity differences
there is no coherence and the value is just the difference between the mean 
value of the transmitted flux and the mean flux in a given flux interval
(Bryan \etal 1998).  It is at intermediate velocity separations where the 
statistic is heavily influenced by the structure of the lines.  There  \LCDM,
OCDM, and sCDM with $\sigma_8=0.7$, whose power spectra are very similar, 
produce very similar distributions; while the tCDM model is quite distinct.
We may quantify these model differences by determining at what $\Delta v$ 
the model prediction passes through a given average flux difference.  For 
the flux interval $0<F<0.1$ the simulation predictions pass through the 
mean flux difference of $0.3$ at $\Delta v \approx 35$~km/s for \LCDM, OCDM, 
and sCDM $\sigma_8=0.7$ and at $\Delta v \approx 45$~km/s for tCDM.  Although 
observational data is limited, these are all lower than the $\Delta v 
\approx 55$~km/s from Figure 3 of Miralde-Escud\'e \etal (1997).

It is important to ask what is needed to restore agreement between the 
simulations and observations.  Although we can not completely rule out the 
possibility that the line fitting algorithm contributes to differences in the
simulated and observed $b$-parameter distributions, we argue that its effect
should not be significant because the discrepancy is seen at a comparable 
level (Bryan \etal 1998) in the fit--independent two--point distribution of 
the flux as well.  The mean optical depth of our models was scaled to agree 
with observations, but this normalization is in some dispute.  However,
changing this normalization has little effect on the median of the 
$b$-distribution. For example, using sCDM ($\sigma_8=0.7$), an increase in 
$\bar\tau_\alpha$ from $0.225$ to $0.35$ at $z=3$ causes the median $b$ 
value to 
decrease from $20.8$~km/s to $20.1$~km/s, a change of only $ \sim 1$~km/s. 
One possibility might be to change the ionization history of the universe 
such that pressure broadening would widen the absorbing structures.  Another 
might be to change the density structure through the power spectrum of the 
cosmology itself.  However, with the suite of models considered here it seems
difficult for a single model to give good agreement with both the column 
density and $b$-parameter data.
 

\section{Flux Statistics for Helium II}
\label{sec:helium}

\begin{figure}
\epsfxsize=3.5in
\centerline{\epsfbox{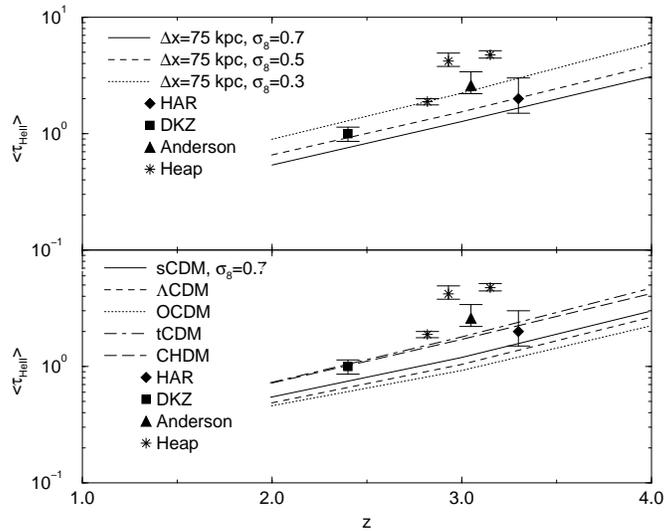}}
\caption{
Evolution of the mean optical depth $\bar \tau_{HeII}$ for \HeII \Lya 
absorption for the sCDM models with $\sigma_8 = 0.7, 0.5$, and $0.3$, 
respectively (top) and for the models of the model comparison study (bottom).
Also shown are observational data with error bars from 
Hogan \etal (HAR, filled diamond), Davidson \etal (DKZ, filled square), 
Anderson \etal (filled triangle), and Heap \etal (stars).} 
\label{fig:Hetau}
\end{figure}

Previous work (Zhang \etal 1997,1998; Croft \etal 1997b) indicates that  
\HeII \Lya (304 \AA) absorption may be significant in regions where 
\HI \Lya is not. Thus the study of \HeII \Lya absorption in quasar spectra 
provides a unique probe of structure in the lowest density regions of the 
universe.  Comparison of both \HI and \HeII absorption within the context of 
a given cosmological model may also yield important information about the 
spectral shape of the metagalactic UV radiation field and its 
redshift evolution. While current observations still struggle to obtain 
sufficient resolution to detect any but the broadest individual \HeII \Lya
lines, it is still possible to determine mean statistics of the \HeII flux 
and optical depth which are not so sensitive to instrumental resolution.
We define the mean optical depth 
$\bar \tau_{HeII} = -\ln \langle F \rangle $ where $\langle F \rangle$ is 
the mean transmitted flux ($F=1$ signifying complete transmission).  
In Figure \ref{fig:Hetau} we present $\bar \tau_{HeII}$ as a function of 
redshift for the sCDM models with varying power normalizations (top) and 
for the models in the model comparison study (bottom).  Several trends are 
apparent. First all models produce a rapid rise in mean optical depth with 
increasing redshift (roughly a factor two between $z=2$ and $z=3$), with 
tCDM and CHDM rising slightly more steeply.  This is consistent with 
previous work on a smaller number of hierarchical cosmologies 
(Zhang \etal 1997; Croft \etal 1997b) and with the interpretation that the 
observed optical depth is due primarily to absorption by gas in underdense 
regions. The redshift evolution of the optical depth is thus dominated by 
the change in the gas density due to universal expansion and 
(to a lesser degree) by the shape of the 
UV metagalactic ionizing background (here assumed to be that of 
Haardt-Madau (1996) with frequency dependence $\propto \nu^{-1.8}$). Second, 
for a given redshift $z$,
models with less power on small fluctuation scales have progressively larger 
optical depths.  This is again consistent with the interpretation that the 
absorption is due to gas in predominantly underdense regions since less gas 
in these low power models will have turned around and collapsed.

The first observation of a flux decrement at the wavelength where the \HeII 
\Lya absorption should occur was made by Jakobson \etal (1994) using the HST 
Faint Object Camera to observe quasar Q0302-003. They  obtained a 
90\% confidence lower limit of $\bar \tau_{HeII}>1.7$ at $z=3.286$.  
Subsequently improved measurements using spectra from this same quasar were 
made by Hogan \etal (1996) with the Goddard High Resolution Spectrograph on 
the HST and by Heap \etal (1998) using the Space Telescope Imaging 
Specrometer (STIS). STIS provides better sensitivity and background 
determinations than previous measurements. 
Davidsen, Kriss \& Zheng (1996) used the Hopkins Ultraviolet Telescope 
to study the average \HeII opacity in the 
spectrum of quasar HS1700+64 over the redshift interval $2.2 < z < 2.6$ 
(lower than that available with HST). They find 
$\bar \tau_{HeII}=1.0\pm0.07$, 
although as shown in Figure 3 of Croft \etal (1997b) there is considerable 
scatter when the wavelength range is divided into $10 \AA$ bins.  
Measurements of \HeII absorption have also been made by Anderson \etal (1998)
with STIS using the spectrum of quasar PKS 1935-692. Although the number of 
lines of sight studied so far are limited and thus a detailed comparison of 
observations with our model simulations (that average over 300 lines of 
sight) is premature, these data are also presented in 
Figure \ref{fig:Hetau}. On face 
value these data favor higher optical depths and thus models with lower 
fluctuation power. However, none of the simulation models presented 
here can reproduce the apparent break at $z=3$ in the optical depth 
observed by Heap \etal (1998).  If this break persists it would most likely 
signal a departure from the Haardt-Madau quasar reionization spectrum assumed 
here.  

\begin{figure}
\epsfxsize=3.5in
\centerline{\epsfbox{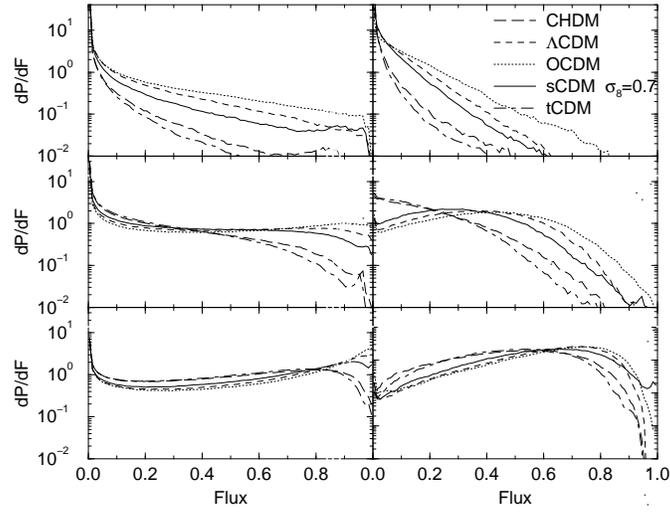}}
\caption{
\HeII \Lya flux probability distribution function smoothed with window 
functions of $50$ \kms (left) and $500$ \kms (right), respectively for 
redshifts (from top to bottom) $z=4,3$, and $2$ for models in the model 
comparison study.}
\label{fig:Hepdf}
\end{figure}

For completeness and comparison with previous work (Croft \etal 1997b)
we show in Figure \ref{fig:Hepdf} (from top to bottom) the \HeII \Lya flux 
probability distribution functions at $z=4,3,2$, respectively, for the 
models in the model 
comparison study.  The distribution functions are calculated from 
the flux smoothed with window functions corresponding to the same Full Width 
at Half Maximum (FWHM) as STIS with high, 50 \kms resolution (left column) 
and low, 500 \kms resolution (right column). As Figure \ref{fig:Hepdf} 
shows, the 
shape of the flux PDF is highly dependent on the smoothing. We see 
however, that models with less fluctuation power on small scales have far 
fewer truly transparent regions (pixels with $F$ near $1$). 
For the high $\Delta v=50$ \kms resolution case, all models converge in the 
fully saturated regime ($F < 0.05$) as expected.


\section{Summary}
\label{sec:summary}

We have performed several simulations of the \Lya forest using
different background cosmological models, numerical codes and
grid resolutions. Five different cosmological models were
considered here: the standard flat critical density cold dark matter model
(sCDM), a flat CDM model with a nonzero cosmological constant (\LCDM),
an open CDM model (OCDM), a flat critical density CDM model with a tilted
power spectrum matching both the COBE amplitude and small scale clustering
contraints (tCDM), and a flat critical density mixed dark matter model (CHDM).
The high resolution shock capturing code Kronos 
was used with grid resolution $\Delta x=37.5$~kpc ($\Delta x=75$~kpc for 
CHDM) for the benchmark calculations presented in this paper.  
Three of the models (sCDM, \LCDM, OCDM with identical parameters) were also 
evolved with the artificial viscosity based code Hercules at the
lower grid resolution $\Delta x = 75$~kpc. Both simulation techniques give 
similar results for statistics, such as the slope of the column density 
distribution, that are insensitive to grid resolution.

We have presented results from several statistical analyses of
absorption features present in the \Lya spectra, both from the unprocessed 
optical depth data and from the reduced line lists.  Explicitly we have 
considered the optical depth and transmitted flux probability distribution 
functions, the cumulative optical depth distributions, the \HI column 
density distributions, line number evolution, Doppler $b$ parameter 
distribution, the average flux difference as a function of velocity (first
moment of the 2-point flux distribution function), and the mean optical depth
and flux probability distribution functions for \HeII absorption.  We find:
\begin{enumerate}
\item Simple scaling laws describe the redshift evolution of the optical 
depth, flux PDF, the \HI column density distribution and, in conjunction 
with the slope of the column density distribution, the line number evolution 
remarkably well.  This demonstrates that most of the evolution of the \Lya 
forest is a direct consequence of universal expansion. 
\item The shape of the optical depth
PDF is strongly correlated to the amplitude of the density fluctuation 
spectrum. Differences between models my be significant in the observationally 
accessible region $0.02 < \tau < 4$. Thus this statistic may be a useful 
discriminator among models.  Similar conclusions hold for the related flux 
PDF. 
\item  Cumulative opacity distributions for the models are strongly clustered
at low optical depth thresholds and high $z$.  Significant differences do 
occur for optical depth thresholds $\tau_0>1$, but these may be more difficult
to observe. 
\item The column density distribution function is a robust statistic
relatively insensitive to grid resolution and numerical method.  Its redshift
evolution is described well by the same naive scaling law that describes the 
evolution of the optical depth.  The slope of the column density distribution
is sensitive to the amplitude of the power spectrum on scales roughly the 
size of the absorbers ($\sim 100$~kpc). Models with less power at these 
scales produce steeper distributions in qualitative agreement with semi-
analytic arguments (Hui, Gnedin \& Zhang 1997). A preliminary comparison with
data favors models with more power (sCDM, \LCDM, OCDM) over those with less 
power (tCDM, CHDM). 
\item All models show comparable evolution for the number
of lines above a given \HI column density threshold in reasonable agreement 
with the data.  Thus this statistic is not a sensitive discriminator among 
models. 
\item Although the shape of the Doppler $b$ parameter distribution is 
well reproduced by all the models, the median of the distribution for sCDM,
\LCDM, and OCDM models is well below observed values.  The median of the $b$ 
parameter increases for models with less power on small scales. Thus the  
observations favor low power models, such as tCDM, making it 
difficult for any model considered in this study to simultaneously give good 
agreement with both the \HI column density and $b$ parameter data. This 
discrepancy is confirmed as well in the nonparametric first moment of the 
two-point flux distribution function and so is not solely the result of the 
line-fitting algorithm employed.  The solution to this problem may require 
modification of the reionization history of the universe to produce more 
pressure broadening of the absorbing structures or a modification of the 
power spectrum of the underlying cosmology itself. 
\item All of the models 
simulated in this study produce a rapid rise in \HeII mean optical depth 
with increasing redshift consistent with the interpretation by previous 
work (Zhang \etal 1997, Croft \etal 1997b) that the observed optical depth 
is due to absorption by gas in underdense regions where universal expansion
dominates the evolution of the gas density.  Models with less power on small
scales (tCDM, CHDM) produce larger mean \HeII optical depths. Preliminary 
comparison with the data tends to favor these low power models. However, none
of the models can reproduce the break seen by Heap \etal near $z=3$. If this 
break persists in the data, it would most likely reflect that the Haardt-
Madau (1996) form for the metagalactic UV ionizing background, based on 
homogeneous reionization by quasars alone in a clumpy medium, must be 
modified.  
\end{enumerate}


\acknowledgements

This work is supported in part by  NSF grant AST-9803137 under the auspices 
of the Grand Challenge Cosmology Consortium (GC$^3$). NASA also supported 
this work through Hubble Fellowship grant HF-0110401-98A from the Space 
Telescope Science Institute, which is operated by the Association of 
Universities for Research in Astronomy, Inc. under NASA contract 
NAS5-26555.  M.M. acknowledges support from the Research and 
Scholarship Development Fund of Northeastern University.  The computations
were performed on the Convex C3880, the SGI Power Challenge,
and the Thinking Machines CM5 at the National Center for Supercomputing 
Applications, and the Cray C90 at the Pittsburgh Supercomputing Center under 
grant AST950004P.



\newpage

\clearpage

\clearpage

\clearpage

\begin{table}
\centerline{\begin{tabular}{|c|c|c|c|c|c|c|c|c|c|c|} \hline
Model & $\Omega_0$ & $\Omega_\Lambda$   & $q_0$ & $\Omega_b$ & 
$h$   & $n$        & $\sigma_{8h^{-1}}$ & $\Delta x$ (kpc) & $\Omega_B h^2$ &  $\sigma_{34}$   \\
\hline 
\hline
sCDM  & 1   & 0   &  0.5   & 0.06   & 0.5  & 1    & 0.7 & 37.5 & $0.015$ & 1.89 \\ \hline
\LCDM & 0.4 & 0.6 & $-0.4$ & 0.0355 & 0.65 & 1    & 1.0 & 37.5 & $0.015$ & 2.03 \\ \hline
OCDM  & 0.4 & 0   &  0.2   & 0.0355 & 0.65 & 1    & 1.0 & 37.5 & $0.015$ & 2.50 \\ \hline
tCDM  & 1   & 0   &  0.5   & 0.07   & 0.6  & 0.81 & 0.5 & 37.5 & $0.025$ & 1.09 \\ \hline
CHDM  & 1   & 0   &  0.5   & 0.07   & 0.6  & 0.98 & 0.7 & 75   & $0.025$ & 1.14 \\ \hline
\end{tabular}}
\caption{Physical parameters of the different cosmological models.
$\Omega_0$ is the total density parameter,
$\Omega_\Lambda = \Lambda/3H_0^2$ the 
cosmological constant density parameter,
$q_0=\Omega_0/2 - \Omega_\Lambda$ the deceleration parameter,
$\Omega_b$ the baryonic mass fraction,
$h$ the Hubble parameter,
$n$ the slope of the primordial density perturbation power spectrum,
$\sigma_{8h^{-1}}$ the fluctuation normalization in a sphere of
radius $8h^{-1}$ Mpc,
$\Delta x$ the comoving simulation spatial resolution in units of kpc, 
$\Omega_b h^2$ is the baryon density in physical units
(independent of $H_0$) and
the last column is proportional to Gnedin's (1998) measure of power at small 
scales. For CHDM the fraction of the energy density carried in 
neutrinos is $0.2$. The sCDM, \LCDM and OCDM models were also simulated with 
the Hercules codes at lower spatial resolution ($\Delta x = 75 $ kpc). 
}
\label{tab:sim_par}
\end{table}

\begin{table}
\centerline{\begin{tabular}{|c|c|c|c|c|c|c|c|} \hline
Model & $z$ & $\beta_\ell$ & $\beta_{\rm HGZ}$ & $\beta_{\rm h}$ & 
$\beta_{\rm f}$ & $\beta_1$ & $\beta_2$  \\
\hline 
\hline
sCDM  & $2$   & $1.64\pm0.03$ & $1.49$ & $1.84\pm0.23$ & $1.87\pm0.05$ & $1.63\pm0.04$ & $1.98\pm0.08$ \\
& $3$   & $1.56\pm0.03$ & $1.55$ & $1.77\pm0.10$ & $1.71\pm0.03$ & $1.52\pm0.04$ & $1.79\pm0.05$ \\ \hline
\LCDM  & $2$  & $1.61\pm0.04$ & $1.50$ & $1.80\pm0.17$ & $1.74\pm0.04$ & $1.56\pm0.05$ & $1.73\pm0.08$ \\
& $3$   & $1.48\pm0.04$ & $1.56$ & $1.78\pm0.12$ & $1.66\pm0.04$ & $1.42\pm0.03$ & $1.74\pm0.03$ \\ \hline
OCDM  & $2$   & $1.56\pm0.04$ & $1.47$ & $1.78\pm0.17$ & $1.71\pm0.04$ & $1.50\pm0.03$ & $1.78\pm0.08$ \\
& $3$   & $1.41\pm0.03$ & $1.52$ & $1.45\pm0.10$ & $1.58\pm0.04$ & $1.41\pm0.05$ & $1.71\pm0.04$ \\ \hline
tCDM  & $2$   & $1.89\pm0.04$ & $1.63$ & $1.93\pm0.18$ & $1.90\pm0.05$ & $1.86\pm0.06$ & $1.83\pm0.13$ \\
& $3$   & $1.72\pm0.08$ & $1.70$ & $2.03\pm0.12$ & $1.95\pm0.06$ & $1.64\pm0.13$ & $2.03\pm0.10$ \\ \hline
CHDM  & $2$  & $1.87\pm0.04$ & $1.66$ & $1.88\pm0.17$ & $1.95\pm0.05$ & $1.85\pm0.05$ & $1.90\pm0.14$ \\
& $3$   & $1.65\pm0.08$ & $1.67$ & $1.89\pm0.10$ & $1.92\pm0.06$ & $1.57\pm0.13$ & $2.01\pm0.08$ \\ \hline
Kim \etal & $2.31$ & $1.35\pm0.03$ & \dots & \dots & \dots & \dots & \dots \\
& $2.85$ & $1.39\pm0.26$ & \dots & $1.46\pm0.07$ & $1.46$ & \dots & \dots \\
& $3.35$ & $1.59\pm0.13$ & \dots & \dots & \dots & \dots & \dots \\ 
& $3.7$  & \dots & \dots & \dots & $1.55$ & \dots & \dots \\ \hline
\end{tabular}}
\caption{Determinations of the \HI column density distribution 
slope over various column density ranges: $\beta_\ell$ over the range 
$10^{12.8}<\NHI<10^{14.3}$ \cms, $\beta_{\rm h}$ over the range 
$10^{13.7}<\NHI<10^{14.3}$ \cms, $\beta_{\rm f}$ over the range 
$10^{12.8}<\NHI<10^{16}$ \cms, $\beta_1$ and $\beta_2$ represent the slopes 
found by splitting the distributions into two halves, 
$10^{12.8}<\NHI<10^{14}$ \cms and $10^{14}<\NHI<10^{16}$ \cms, respectively. 
The column labeled $\beta_{\rm HGZ}$ is the prediction of the semi-analytic
model of Hui \etal (1997). The last row provides the measured values
reported by Kim \etal (1997).  All quoted errors are $2\sigma$. 
}
\label{tab:beta}
\end{table}

\begin{table}
\centerline{\begin{tabular}{|c|c|c|c|c|c|} \hline
Model & $\gamma_{13}$ & $\gamma_{13.5}$ & $\gamma_{14}$ & $\gamma_{\rm h}$ & 
$\gamma_\ell$ \\ 
\hline
\hline
sCDM & $2.48\pm0.13$ & $3.49\pm1.10$ & $4.27\pm1.65$ & $2.97$ & $2.29$ \\ 
\hline
\LCDM & $2.42\pm0.36$ & $3.38\pm0.92$ & $4.09\pm1.42$ & $2.91$ & $2.02$ \\ 
\hline
OCDM & $2.18\pm0.89$ & $2.96\pm0.78$ & $3.71\pm0.95$ & $2.73$ & $1.72$ \\ 
\hline
tCDM & $2.60\pm0.79$ & $4.29\pm1.10$ & $5.30\pm2.79$ & $3.49$ & $3.28$ \\ 
\hline
CHDM & $2.32\pm0.97$ & $4.09\pm0.62$ & $5.36\pm2.25$ & $3.79$ & $3.10$ \\ 
\hline
combined & $1.61\pm0.66$ & $2.82\pm0.83$ & $3.47\pm1.23$ & \dots & \dots \\ 
\hline
Kim \etal & \dots & \dots & \dots & $2.78\pm1.42$ & $1.29\pm0.29$ \\ \hline
\end{tabular}}
\caption{Slope of the predicted line number evolution  for models in the 
model comparison study as a function of redshift  fit to the 
form $dN/dz \propto (1+z)^\gamma$  for various fixed column density 
thresholds: $\NHI > 10^{13}$  ($\gamma_{13}$), $\NHI>10^{13.5}$ 
($\gamma_{13.5}$), $\NHI>10^{14}$  ($\gamma_{14}$), for $2\le z\le 4$ and 
column density ranges: $10^{13.77}<\NHI<10^{16}$ ($\gamma_{\rm h}$), 
$10^{13.1}<\NHI<10^{14}$ ($\gamma_\ell$) for $z=2$ and $3$. The same fits are
presented for the combined line lists of Kulkarni \etal, Hu \etal,
Kirkman \& Tytler, and Lu \etal (combined) and for data from Kim \etal 
(1997). All quoted errors are $2\sigma$.
}
\label{tab:gamma}
\end{table}

\begin{thebibliography}{}

\bibitem[]{aazn97}
        Abel, T., Anninos, P., Zhang, Y., \& Norman, M. L.
        1997, {New Astronomy}, 2, 181

\bibitem[]{and98}
        Anderson, S.~F., Hogan, C.~J., Williams, B.~F., \& Carswell, R.~F. 
        1998, \aj, in press (astro-ph/9808105)

\bibitem[]{ann94}
        Anninos, P., Norman, M. L., \& Clarke, D. A. 1994, \apj, 436, 11

\bibitem[]{azan97}
        Anninos, P., Zhang, Y., Abel, T., \& Norman, M. L.
        1997, {New Astronomy}, 2, 209

\bibitem[]{BBKS96}
        Bardeen, J.~M., Bond, J.~R., Kaiser, N., \& Szalay, A.~S. 1996,
        \apj, 304, 15

\bibitem[]{ber95}
        Bertschinger, E. 1995, astro-ph/9506070

\bibitem[]{bon96}
        Bond, J.~R., \& Myers, S.~T. 1996, \apjs, 103, 63

\bibitem[]{bon97}
        Bond, J.~R., \& Wadsley, J.~W. 1997, astro-ph/9710102

\bibitem[]{bry95}
        Bryan, G.~L., Norman, M.~L., Stone, J.~M., Cen, R., Ostriker, J.P.
        1995, {Comput. Phys. Comm.}, 89, 149

\bibitem[]{bry98}
        Bryan, G.~L., Machacek, M., Anninos, P., \& Norman, M.~L. 1998, \apj,
        in press (astro-ph/9805340)

\bibitem[]{bun97}
        Bunn, E~F., \& White, M. 1997, \apj, 480, 6

\bibitem[]{bur98}
        Burles, S., \& Tytler, D. 1998, \apj, 499, 699

\bibitem[]{car87}
        Carswell, R.~F., Webb, J.~K., Baldwin, J.~A., \& Atwood, B. 1987,
        \apj, 319, 709

\bibitem[]{cen96}
        Cen, R. 1997, \apj, 479, L85

\bibitem[]{cen94}
        Cen, R., Miralda-Escud\'e,J., Ostriker, J. P., \& Rauch, M.
        1994, \apj, 437, L9

\bibitem[]{cha96}
        Charlton, J., Anninos, P., Zhang, Y., \& Norman, M. L.
        1997, \apj, 485, 26

\bibitem[]{cop95}
        Copi, C. J., Schramm, D. N., \& Turner, M. S. 1995, Science, 267, 192

\bibitem[]{croft97a}
        Croft, R.~A.~C., Weinberg, D.~H., Hernquist, L., Katz, N. 1997a, 
        in the proceedings of the 18th Texas Symposium on Relativistic
        Astrophysics, Chicago, December 1996, eds A. Olinto, J. Frieman and 
        D. Schramm (World Scientific)

\bibitem[]{croft97b}
        Croft, R.~A.~C., Weinberg, D.~H., Katz, N., \& Hernquist, L. 1997b,
        \apj, 488, 532

\bibitem[]{croft98}
        Croft, R.~A.~C., Weinberg, D.~H., Pettini, M., Hernquist, L., \& 
        Katz, N. 1998, \apj, submitted (astro-ph/9809401) 
 
\bibitem[]{dav97}
        Dav\'e, R., Hernquist, L., Weinberg, D.~H., \& Katz, N. 1997, \apj,
        477, 21

\bibitem[]{dav96}
        Davidsen, A., Kriss, G. A., \& Zheng, W. 1996, Nature, 380, 47

\bibitem[]{EH98}
        Eisenstein, D.~J., \& Hu, W. 1998, \apj, 496, 605

\bibitem[]{gia96}
        Giallongo, E., Cristiani, S., D'Odorico, S., Fontana, A., \&
        Savaglio, S. 1996, \apj, 466, 46

\bibitem[]{gned98}
        Gnedin, N.~Y. 1998, MNRAS, submitted (astro-ph/9706286)


\bibitem[]{haa96}
        Haardt, F. \& Madau, P. 1996, \apj, 461, 20

\bibitem[]{heap}
        Heap, S.~R., Williger, G.~M., Smette, A., Hubeny, I., Sahu, M., 
        Jenkins, E.~B., Tripp, T.~M., \& Winkler, J.~N., 1998, prepint
        (astro-ph/9812429)

\bibitem[]{her96}
        Hernquist, L., Katz, N., Weinberg, D., \& Miralda-Escud\'e, J.
        1996, \apj, 457, L51

\bibitem[]{hogan}
        Hogan, C., Anderson, S., \& Rugers, M. 1997, \aj, 113, 1495

\bibitem[]{hu95}
        Hu, E. M., Kim, T. S., Cowie, L. L., Songaila, A., \& Rauch, M.
        1995, \aj, 110, 1526

\bibitem[]{HGZ97}
        Hui, L., Gnedin, N.~Y., \& Zhang, Y. 1997, \apj, 486, 599

\bibitem[]{huirut}
        Hui, L. \& Rutledge, R.E. 1998, preprint (astro-ph/9709100)

\bibitem[]{jak94}
        Jakobsen, P., Boksenberg, A., Deharveng, J. M., Greenfield, P.,
        Jedrzejewski, R., \& Paresce, F. 1994, Nature, 370, 35

\bibitem[]{ken95}
        Kennefick, J.D. \etal 1995, \aj, 110, 78

\bibitem[]{kim97}
        Kim, T.-S., Hu, E.~M., Cowie, L.~L., \& Songaila, A. 1997, \aj, 114, 1

\bibitem[]{kir97}
        Kirkman, D., \& Tytler, D. 1997, \apj, 484, 672

\bibitem[]{kulk96}
        Kulkarni, V.P., Huang, K., Green, R.F., Bechtold, J.,
        Welty, D.E. \& York, D.G. 1996, MNRAS, 279, 197

\bibitem[]{lu97}
        Lu, L., Sargent, W.~L.~W., Womble, D.~S., \& Takada-Hidai, M. 1997,
        \apj, 472, 509

\bibitem[]{ma96}
        Ma, C.-P. 1996, \apj, 471, 13

\bibitem[]{mr93}
        Maoz, D. \& Rix, H.W. 1993, \aj, 416, 425

\bibitem[]{mei98}
        Meiksin, A.  \etal 1999, in preparation

\bibitem[]{mir96}
        Miralda-Escud\'e, J., Cen, R., Ostriker, J. P., \& Rauch, M.
        1996, \apj, 471, 582

\bibitem[]{mir97}
        Miralda-Escud\'e, J., Rauch, M., Sargent, W.~L.~W., Barlow, T.~A., 
        Weinberg, D.~H., Hernquist, L., Katz, N., Cen, R., Ostriker, J.~P. 
        1997, To appear in Proceedings of 13th IAP Colloquium: Structure and
        Evolution of the IGM from QSO Absorption Line Systems, eds. P. 
        Petijean, S. Charlot

\bibitem[]{oke82}
        Oke, J.B., \& Korycansky, D.G. 1982, \apj, 255, 11

\bibitem[]{ost95}
        Ostriker, J.~P., \& Steinhardt, P.~J. 1995, Nature, 377, 600

\bibitem[]{pet93}
        Petitjean, P., Webb, J. K., Rauch, M., Carswell, R. F., \& Lanzetta, 
        K. 1993, \mnras, 262, 499

\bibitem[]{PRS93}
        Press, W.~H., Rybicki, G.~B., \& Schneider, D.~P. 1993, \apj, 414, 64

\bibitem[]{Rau97}
        Rauch, M. \etal 1997, \apj, 489, 7

\bibitem[]{Rut98}
        Rutledge, R.L. 1998, \apj, submitted (astro-ph/9707334)

\bibitem[]{sch91}
        Schneider, D.P., Schmidt, M., \& Gunn, J.E. 1991, \aj, 101, 2004

\bibitem[]{sel96}
        Seljak, U. \& Zaldarriaga, M. 1996, \apj, 469, 437


\bibitem[]{ste87}
        Steidel, C.C., \& Sargent, W.L.W. 1987, \apj, 313, 171

\bibitem[]{the98}
        Theuns, T., Leonard, A., \& Efstathiou, G. 1998, \mnras, 297, L49

\bibitem[]{whi93}
        White, S.~D.~M., Efstathiou, G., \& Frenk, C.~S. 1993, \mnras, 262, 
        1023

\bibitem[]{wein}
        Weinberg, D.H., Hernquist, L., Katz, N., \& Miralda-Escud\'e, J., 
        In  Cold Gas at High Redshift, eds. Bremer, M., Rottgering, H., 
        Carilli, C., \& van der Werf, P., Kluwer: Dordrecht

\bibitem[]{zha95}
        Zhang, Y., Anninos, P., \& Norman, M. L. 1995, \apj, 453, L57

\bibitem[]{zman97}
        Zhang, Y., Anninos, P., Norman, M. L. \& Meiksin, A.
        1997, \apj, 485, 496

\bibitem[]{zman98}
        Zhang, Y., Meiksin, A., Anninos, P., \& Norman, M. L.
        1998, \apj, 495, 63

\bibitem[]{zuo93}
        Zuo, L., \& Lu, L. 1993, \apj, 418, 601

\end{thebibliography}
\end{document}